\documentclass[12pt,a4paper]{article}
\usepackage{ifthen} % for conditional statements
\usepackage{feynmp}
\usepackage{subfigure}

\usepackage{hepnicenames}

\usepackage{pbox}
\usepackage{color}
\usepackage[usenames,dvipsnames]{xcolor}

\newboolean{pdflatex}
\setboolean{pdflatex}{true} % False for eps figures 

\newboolean{articletitles}
\setboolean{articletitles}{true} % False removes titles in references

\newboolean{uprightparticles}
\setboolean{uprightparticles}{false} %True for upright particle symbols

\newboolean{inbibliography}
\setboolean{inbibliography}{false} %True once you enter the bibliography

\usepackage[top=1in, bottom=1.25in, left=1in, right=1in]{geometry}

\usepackage[percent]{overpic}
\usepackage[normalem]{ulem}

\usepackage{microtype}
\usepackage{lineno}  % for line numbering during review
\usepackage{xspace} % To avoid problems with missing or double spaces after
\usepackage{caption} %these three command get the figure and table captions automatically small

\usepackage{graphicx}  % to include figures (can also use other packages)
\usepackage{color}
\usepackage{colortbl}
\graphicspath{{./figs/}} % Make Latex search fig subdir for figures

\usepackage{amsmath} % Adds a large collection of math symbols
\usepackage{amssymb}
\usepackage{amsfonts}
\usepackage{upgreek} % Adds in support for greek letters in roman typeset

\newcommand*\patchAmsMathEnvironmentForLineno[1]{%
\expandafter\let\csname old#1\expandafter\endcsname\csname #1\endcsname
\expandafter\let\csname oldend#1\expandafter\endcsname\csname
end#1\endcsname
 \renewenvironment{#1}%
   {\linenomath\csname old#1\endcsname}%
   {\csname oldend#1\endcsname\endlinenomath}%
}
\newcommand*\patchBothAmsMathEnvironmentsForLineno[1]{%
  \patchAmsMathEnvironmentForLineno{#1}%
  \patchAmsMathEnvironmentForLineno{#1*}%
}
\AtBeginDocument{%
\patchBothAmsMathEnvironmentsForLineno{equation}%
\patchBothAmsMathEnvironmentsForLineno{align}%
\patchBothAmsMathEnvironmentsForLineno{flalign}%
\patchBothAmsMathEnvironmentsForLineno{alignat}%
\patchBothAmsMathEnvironmentsForLineno{gather}%
\patchBothAmsMathEnvironmentsForLineno{multline}%
\patchBothAmsMathEnvironmentsForLineno{eqnarray}%
}

\usepackage{hyperref}    % Hyperlinks in references
\usepackage[all]{hypcap} % Internal hyperlinks to floats.

\def\xsec {cross-section\xspace}

\def\ptWZ  {\ensuremath{p^{\Z/\W}_{\text{T}}}\xspace}
\def\ptZ  {\ensuremath{p^{\Z}_{\text{T}}}\xspace}
\def\ptl  {\ensuremath{p^{l}_{\text{T}}}\xspace}

\def\ptWZsq  {\ensuremath{(p^{\Z/\W}_{\text{T}})^2}\xspace}
\def\njet  {\ensuremath{N_{\text{jet}}}\xspace}
\def\jetpt  {\ensuremath{p^{\text{jet}}_{\text{T}}}\xspace}

\def\kt  {\ensuremath{k_{\text{T}}}\xspace} 

\def\lhcb {\mbox{LHCb}\xspace}

\def\MagUp {\mbox{\em Mag\kern -0.05em Up}\xspace}

\ifthenelse{\boolean{uprightparticles}}%
{

 \def\PDelta      {\ensuremath{\Delta}\xspace}                 
 \def\PXi      {\ensuremath{\Xi}\xspace}                 
 \def\PLambda      {\ensuremath{\Lambda}\xspace}                 
 \def\PSigma      {\ensuremath{\Sigma}\xspace}                 
 \def\POmega      {\ensuremath{\Omega}\xspace}                 
 \def\PUpsilon      {\ensuremath{\Upsilon}\xspace}                 
 
 \def\PB      {\ensuremath{\mathrm{B}}\xspace}                 
                  
 \def\PD      {\ensuremath{\mathrm{D}}\xspace}

 \def\PK      {\ensuremath{\mathrm{K}}\xspace}

 \def\PW      {\ensuremath{\mathrm{W}}\xspace}

 \def\PZ      {\ensuremath{\mathrm{Z}}\xspace}

 \def\Pi      {\ensuremath{\mathrm{i}}\xspace}

}
{

 \mathchardef\PDelta="7101
 \mathchardef\PXi="7104
 \mathchardef\PLambda="7103
 \mathchardef\PSigma="7106
 \mathchardef\POmega="710A
 \mathchardef\PUpsilon="7107
                  
 \def\PB      {\ensuremath{B}\xspace}                 
                  
 \def\PD      {\ensuremath{D}\xspace}

 \def\PK      {\ensuremath{K}\xspace}

 \def\PW      {\ensuremath{W}\xspace}

 \def\PZ      {\ensuremath{Z}\xspace}

 \def\Pi      {\ensuremath{i}\xspace}

}

\makeatletter
\ifcase \@ptsize \relax% 10pt
  \newcommand{\miniscule}{\@setfontsize\miniscule{4}{5}}% \tiny: 5/6
\or% 11pt
  \newcommand{\miniscule}{\@setfontsize\miniscule{5}{6}}% \tiny: 6/7
\or% 12pt
  \newcommand{\miniscule}{\@setfontsize\miniscule{5}{6}}% \tiny: 6/7
\fi
\makeatother

\DeclareRobustCommand{\optbar}[1]{\shortstack{{\miniscule (\rule[.5ex]{1.25em}{.18mm})}
  \\ [-.7ex] $#1$}}

\def\W      {{\ensuremath{\PW}}\xspace}

\def\Z      {{\ensuremath{\PZ}}\xspace}

  \def\Kbar    {{\kern 0.2em\overline{\kern -0.2em \PK}{}}\xspace}

\def\KorKbar    {\kern 0.18em\optbar{\kern -0.18em K}{}\xspace}

  \def\Dbar    {{\kern 0.2em\overline{\kern -0.2em \PD}{}}\xspace}

\def\DorDbar    {\kern 0.18em\optbar{\kern -0.18em D}{}\xspace}

\def\Bbar    {{\ensuremath{\kern 0.18em\overline{\kern -0.18em \PB}{}}}\xspace}

\def\BorBbar    {\kern 0.18em\optbar{\kern -0.18em B}{}\xspace}

  \def\Y#1S{\ensuremath{\PUpsilon{(#1S)}}\xspace}% no space before {...}!

\def\Lbar        {{\ensuremath{\kern 0.1em\overline{\kern -0.1em\PLambda}}}\xspace}
\def\LorLbar    {\kern 0.18em\optbar{\kern -0.18em \PLambda}{}\xspace}

\def\to                 {\ensuremath{\rightarrow}\xspace}

\newcommand{\as}{{\ensuremath{\alpha_s}}\xspace}

\def\AT#1     {\ensuremath{A_{\mathrm{T}}^{#1}}\xspace}           % 2

\def\C#1      {\ensuremath{\mathcal{C}_{#1}}\xspace}                       % 9
\def\Cp#1     {\ensuremath{\mathcal{C}_{#1}^{'}}\xspace}                    % 7
\def\Ceff#1   {\ensuremath{\mathcal{C}_{#1}^{\mathrm{(eff)}}}\xspace}        % 9  
\def\Cpeff#1  {\ensuremath{\mathcal{C}_{#1}^{'\mathrm{(eff)}}}\xspace}       % 7
\def\Ope#1    {\ensuremath{\mathcal{O}_{#1}}\xspace}                       % 2
\def\Opep#1   {\ensuremath{\mathcal{O}_{#1}^{'}}\xspace}                    % 7

\newcommand{\tev}{\ifthenelse{\boolean{inbibliography}}{\ensuremath{~T\kern -0.05em eV}\xspace}{\ensuremath{\mathrm{\,Te\kern -0.1em V}}}\xspace}
\newcommand{\gev}{\ensuremath{\mathrm{\,Ge\kern -0.1em V}}\xspace}
\newcommand{\mev}{\ensuremath{\mathrm{\,Me\kern -0.1em V}}\xspace}
\newcommand{\kev}{\ensuremath{\mathrm{\,ke\kern -0.1em V}}\xspace}
\newcommand{\ev}{\ensuremath{\mathrm{\,e\kern -0.1em V}}\xspace}
\newcommand{\gevc}{\ensuremath{{\mathrm{\,Ge\kern -0.1em V\!/}c}}\xspace}
\newcommand{\mevc}{\ensuremath{{\mathrm{\,Me\kern -0.1em V\!/}c}}\xspace}
\newcommand{\gevcc}{\ensuremath{{\mathrm{\,Ge\kern -0.1em V\!/}c^2}}\xspace}
\newcommand{\gevgevcccc}{\ensuremath{{\mathrm{\,Ge\kern -0.1em V^2\!/}c^4}}\xspace}
\newcommand{\mevcc}{\ensuremath{{\mathrm{\,Me\kern -0.1em V\!/}c^2}}\xspace}

\def\gsim{{~\raise.15em\hbox{$>$}\kern-.85em
          \lower.35em\hbox{$\sim$}~}\xspace}
\def\lsim{{~\raise.15em\hbox{$<$}\kern-.85em
          \lower.35em\hbox{$\sim$}~}\xspace}

\def\pt         {\mbox{$p_{\rm T}$}\xspace}

\def\pythia     {\mbox{\textsc{PYTHIA}}\xspace}
\def\rivet     {\mbox{\textsc{RIVET}}\xspace}
\def\resbos     {\mbox{\textsc{ResBos}}\xspace}

\def\mnm     {\mbox{\textsc{M\&M}}\xspace}
\def\ckkw     {\mbox{\textsc{CKKW}}\xspace}
\def\ckkwl     {\mbox{\textsc{CKKW-L}}\xspace}
\def\mlm     {\mbox{\textsc{MLM}}\xspace}
\def\umeps     {\mbox{\textsc{UMEPS}}\xspace}
\def\fxfx    {\mbox{\textsc{FxFx}}\xspace}
\def\unlops     {\mbox{\textsc{UNLOPS}}\xspace}
\def\unnlops     {\mbox{\textsc{U$\text{N}^2$LOPS}}\xspace}
\def\mcatnlo    {\mbox{\textsc{MC@NLO}}\xspace}
\def\amcatnlo    {\mbox{a\textsc{MC@NLO}}\xspace}
\def\madgraph    {\mbox{\textsc{MadGraph}}\xspace}
\def\nl3     {\mbox{\textsc{N$\text{L}^3$}}\xspace}

\def\tell1  {TELL1\xspace}
\def\ukl1   {UKL1\xspace}

\newcommand{\ie}{\mbox{\itshape i.e.}\xspace}

\usepackage{cite} % Allows for ranges in citations
\usepackage{mciteplus}

\usepackage{longtable} % only for template; not usually to be used in PAPERs

\begin{document}

\begin{fmffile}{fgraphs}

\renewcommand{\thefootnote}{\fnsymbol{footnote}}
\setcounter{footnote}{1}

%\onecolumn
\begin{titlepage}

\vspace*{-1.5cm}

\hspace*{-0.5cm}
\begin{tabular*}{\linewidth}{lc@{\extracolsep{\fill}}r}
%\ifthenelse{\boolean{pdflatex}}% Logo format choice
%{\vspace*{-2.7cm}\mbox{\!\!\!\includegraphics[width=.14\textwidth]{lhcb-logo.pdf}} & &}%
%{\vspace*{-1.2cm}\mbox{\!\!\!\includegraphics[width=.12\textwidth]{lhcb-logo.eps}} & &}
 \\
 & & MCnet-16-37 \\  % ID 
 & & August, 2016 \\ % Date - Can also hardwire e.g.: 23 March 2010
 & & \\
\hline
\end{tabular*}

\vspace*{4.0cm}

% Title --------------------------------------------------
{\bf\boldmath\huge
\begin{center}
  Modelling electroweak physics\linebreak for the forward region
\end{center}
}

\vspace*{2.0cm}

% Authors -------------------------------------------------
\begin{center}
M.~Sirendi$^{1,2}$
\bigskip\\
{\it\footnotesize
$ ^1$Cavendish Laboratory, University of Cambridge, United Kingdom\\
$ ^2$Dept. of Astronomy and Theoretical Physics, Lund University, Sweden\\
}
\end{center}

\vspace{\fill}

% Abstract -----------------------------------------------
\begin{abstract}
  \noindent
  This note presents a study of matching and merging schemes and weak showering at the LHC as applied to the production of electroweak bosons in association with jets. These advanced theoretical tools are seen to provide a good description of event shapes in the central region when compared to measurements performed by the ATLAS and CMS collaborations at a centre-of-mass energy of 7\tev. Matching and merging schemes also provide a superior description of forward \mbox{$\Z+$jets} production. The study constitutes a test of matching and merging schemes in a novel region of phase space and can be considered as a validation of the universality of these techniques. Finally, it was determined that current measurements of electroweak physics at LHCb do not yet fully probe the effect of weak showers.
\end{abstract}

\vspace*{2.0cm}
\vspace{\fill}

\end{titlepage}

\pagestyle{empty}  

\newpage
\setcounter{page}{2}
\mbox{~}

\cleardoublepage
%\twocolumn

\renewcommand{\thefootnote}{\arabic{footnote}}
\setcounter{footnote}{0}

\tableofcontents
\newpage
%\cleardoublepage

\pagestyle{plain} % restore page numbers for the main text
\setcounter{page}{1}
\pagenumbering{arabic}

%\linenumbers

\section{Introduction}
\label{sec:Introduction}

The study of $\Z/\W+$jets production\footnote{Throughout this note \Z denotes the combined \Z and virtual photon ($\gamma^*$) contribution.} in $pp$ collisions serves as a testing ground for theoretical predictions derived from perturbative quantum chromodynamics (pQCD) and electroweak (EW) theory. These processes are also a ubiquitous background to physics beyond the Standard Model (SM) and so must be precisely measured and theoretically understood. Measurements performed by the ATLAS~\cite{WZATLAS, ATLASDY8TeV, ATLAS13TeV}, CMS~\cite{CMSWZpT8TeV, CMSWasymmetry8TeV, CMSWZ8TeV}, and LHCb~\cite{WenuLHCb8TeV, WmunuLHCb, WZLHCb8TeV, ZmumuLHCb, ZeeLHCb, ZeeLHCb8TeV, ZtautauLHCb, Z13TeV} collaborations are in good agreement with theoretical predictions that are determined from parton-parton cross-sections convolved with parton distribution functions (PDFs). The precision of these predictions is limited by the accuracy of the PDFs and by higher-order QCD corrections which are currently known at next-to-next-to-leading order (NNLO) in pQCD~\cite{FEWZ,FEWZ2}.

Two conventional theoretical tools employed in high energy physics to make QCD predictions are fixed order calculations and Monte Carlo (MC) event generators. Fixed order calculations are limited in their ability to describe event topologies with many jets. Conversely, MC event generators commonly employ the parton shower (PS) formalism which approximates parton emissions to all orders and multiplicities. This approach, however, mismodels the number, and hardness, of jets due to its approximate nature.

This note concerns itself with the study of two novel theoretical approaches for rectifying the above issues: matching and merging (\mnm) and weak showering (WS). \mnm seeks to unite a matrix element (ME) calculation at a given order in pQCD with the PS. On the other hand, WS improves the PS description of jet formation by allowing for the emission of EW bosons within the jet. In particular, this note will compare these theoretical approaches to measurements performed by the \lhcb experiment in the forward region at the LHC. The LHCb detector, which is instrumented in the pseudorapidity range $2.0<\eta<5.0$, is in a unique situation to validate the universality of these techniques and act as a complementary test to measurements performed by the ATLAS and CMS collaborations.

There are several other theoretical effects that affect the performance of these predictions such as higher order quantum electrodynamic (QED) and EW corrections, their interplay and effect on pQCD corrections. Within the realm of MC event generators, the tuning of the PS and multiparton interactions (MPI) becomes important. Finally, uncertainties due to PDFs enter and can become particularly large at high and low values of Bjorken-$x$ which corresponds to the kinematic acceptance of the \lhcb detector. While this note seeks to contrast different \mnm schemes and test the WS formalism, other studies have devoted considerable effort in understanding the other effects mentioned~\cite{PDF4LHC,Alioli}.

In this note, we present an overview of fixed order calculations and the factorisation theorem in Sect.~\ref{sec:hadroproductionWeakGaugeBosons}, an introduction to the PS formalism in Sect.~\ref{sec:partonShowerFramework}, a comparison of \mnm schemes in Sect.~\ref{sec:matchingMerging} and an explanation of WS in Sect.~\ref{sec:weakGaugeBoson}. We compare \mnm schemes and WS to measurements performed by ATLAS~\cite{ATLASZjets}, CMS~\cite{CMSZeventShapes, CMSWjets}, and \lhcb~\cite{LHCbZjets} at a centre-of-mass energy of $\sqrt{s}=7$\tev. We also compare \mnm schemes to a measurement of inclusive \Z production by \lhcb at $\sqrt{s}=13$\tev~\cite{Z13TeV}.

\section{The hadroproduction of weak gauge bosons}
\label{sec:hadroproductionWeakGaugeBosons}

\begin{figure}[tb]
\begin{center}
	\includegraphics[width=0.49\linewidth]{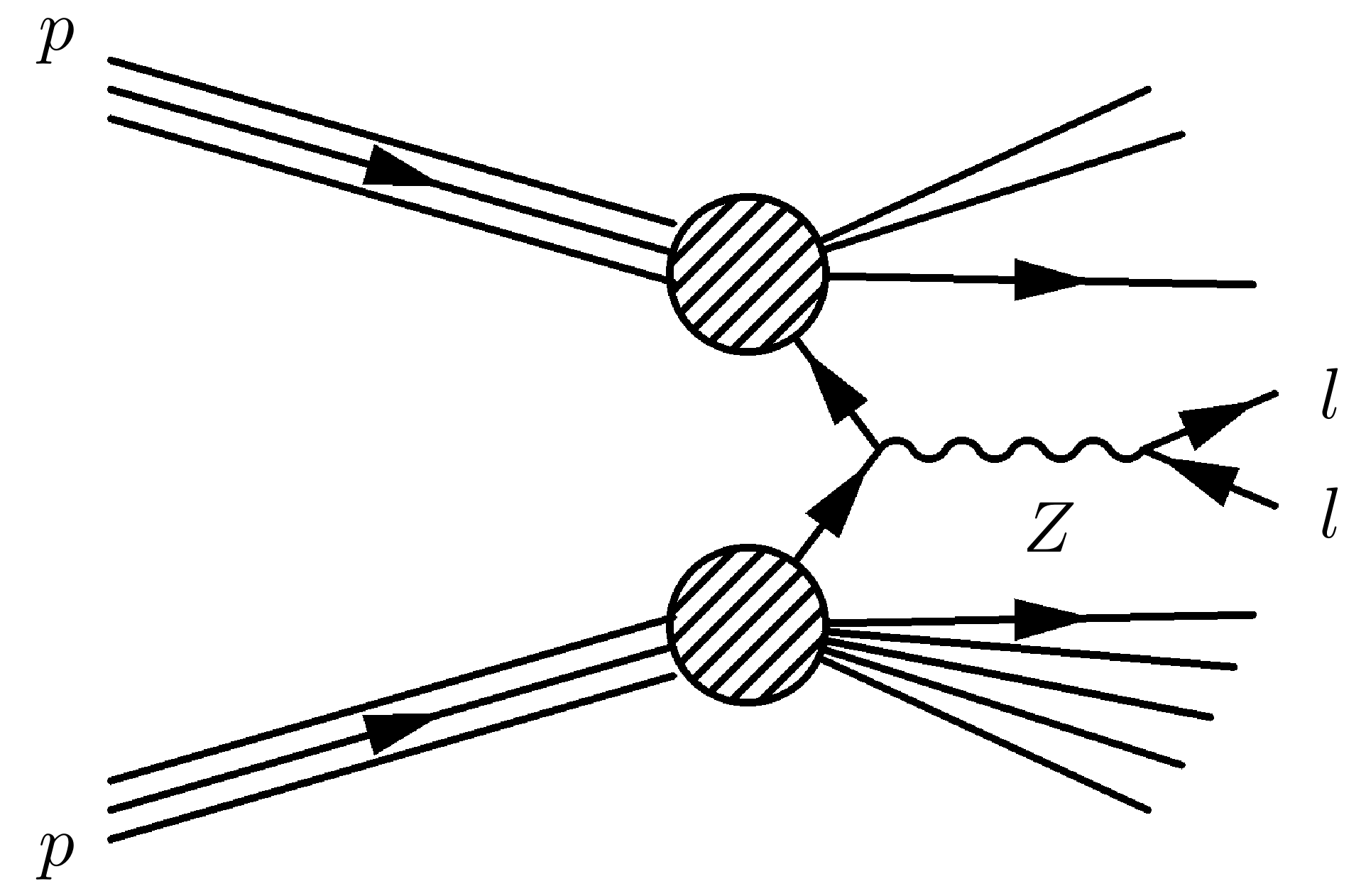}
\vspace*{-0.5cm}
\end{center}
\caption{
	The Drell-Yan process for producing a \Z boson in a $pp$ collision.}
\label{fig:drellYan}
\end{figure}

Drell and Yan first proposed that a scattering \xsec in hadronic collisions can be treated as a convolution of the partonic \xsec, calculable in pQCD, and the distributions of parton momentum in the proton, given by a PDF~\cite{drellYan}. The archetypal example for the procedure is provided by the Drell-Yan interaction, illustrated in Fig.~\ref{fig:drellYan}. However, it has been successfully extended to other processes and, crucially, it enables higher order QCD corrections to be incorporated. In general, a \xsec, $\sigma_{pp\to X}$, summed over initial state partons $a$ and $b$ with momentum fractions $x_a$ and $x_b$, with PDFs $f_a(x_a)$ and $f_b(x_b)$, and at a given energy scale $Q$ is

\begin{equation}
\label{eq:factTheorem}
	\sigma_{pp\to X}=\textup{PDF}\otimes\hat{\sigma}_{ab\to X}=\sum_{a,b}{\int{dx_a\,dx_b\,f_a(x_a,Q^2)\,f_b(x_b,Q^2)\,\hat{\sigma}_{ab\to X}(Q^2)}}.
\end{equation}

The partonic \xsec, $\hat{\sigma}_{ab\to X}$, is calculated as a power series expansion in the strong coupling constant, \as, with further terms corresponding to higher order emissions:

\begin{equation}
\label{eq:sigmaExpansion}
	\hat{\sigma}_{ab\to X}=[\hat{\sigma}_{\textup{LO}}+\as \hat{\sigma}_{\textup{NLO}}+\alpha_s^2\hat{\sigma}_{\textup{NNLO}}+...]_{ab\to X}.
\end{equation}

\begin{figure}
\begin{center}
	\includegraphics[width=0.8\linewidth]{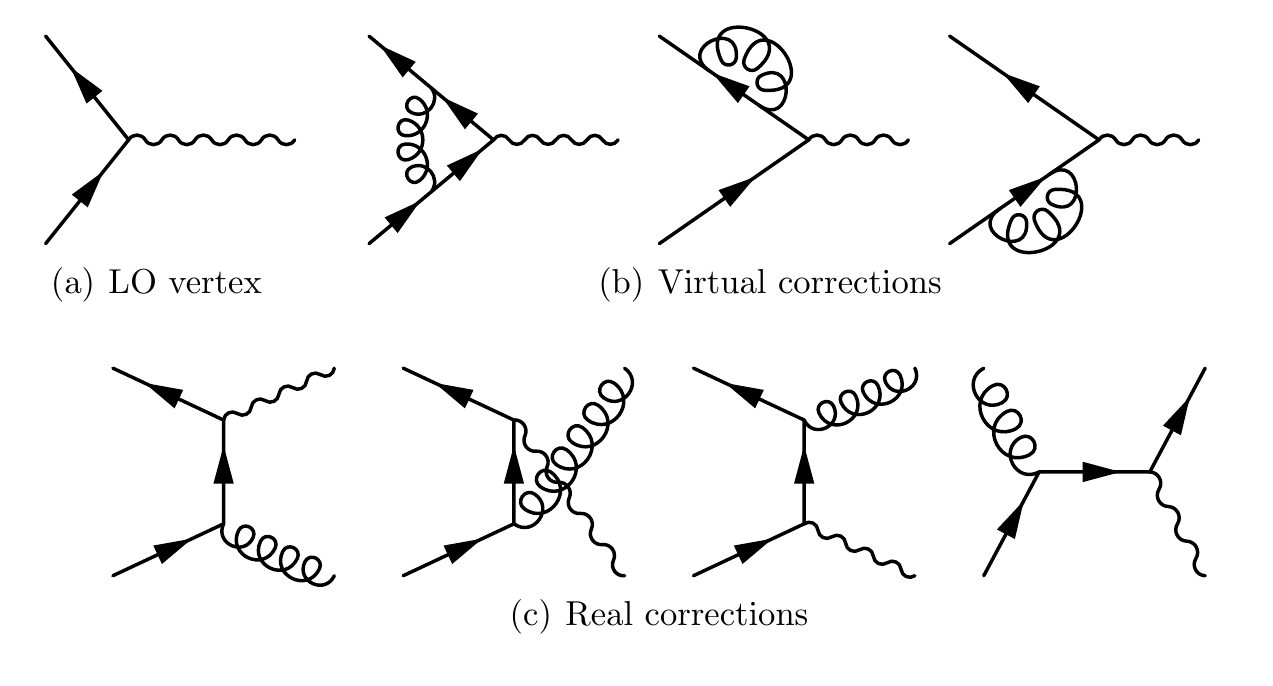}
\end{center}
\vspace{-0.5cm}
\caption{The LO interaction vertex with first order corrections.}
\label{fig:drellYanFD}
\end{figure}

\noindent Feynman diagrams for leading virtual and real corrections (\ie NLO) are illustrated in Fig.~\ref{fig:drellYanFD}. The higher order terms increase the LO \xsec for \W and \Z bosons by about $20-30\%$~\cite{qcdAndCollider} and are needed to reproduce the boson transverse momentum, \ptWZ, spectrum observed in data. Generally, the precision of the measurement dictates the order of the perturbative calculation that must be performed to make a valid comparison.

For the Drell-Yan process specifically terms up to $\mathcal{O}(\alpha_s^2)$ have been computed~\cite{drellYanNNLO} with recent progress on the $\mbox{N}^3$LO calculation~\cite{drellYanN3LO}. However, as the number of Feynman diagrams increases roughly factorially with each order in perturbation theory, it becomes increasingly difficult to extend the precision of these calculations. What is more, at this order in \as QED and EW corrections become comparable in size and must also be computed.

There is also an additional difficulty with this approach. Namely, when the mass of the boson, $M_{\Z/\W}$, is significantly greater than its \ptWZ (\ie close to threshold production), higher-order terms become large and spoil the convergence of the series. In particular, the emission of multiple soft gluons must be considered. The leading contributions at each order are proportional to

\begin{equation}
\label{logs}
	\as \ln{\frac{M^2_{\Z/\W}}{\ptWZsq}}.
\end{equation}

\noindent Clearly these logarithms are significant when \ptWZ is small. In fact, after taking prefactors into account, these terms become $\mathcal{O}(1)$ for \ptWZ values\footnote{Natural units with $\hbar=c=1$ are used throughout.} less than \mbox{$10-15\gev$~\cite{qcdAndCollider}}. Fortunately, these terms can be resummed to all orders. This is particularly advantageous near the boundaries of phase space where fixed order calculations break down. The method of resumming logarithms has been implemented in the \resbos~\cite{RESBOS1,RESBOS2,RESBOS3} event generator where next-to-next-to-leading log (NNLL) corrections are computed for NLO differential cross-sections.

The inclusive \Z or \W cross-section can also be decomposed into its multijet components, formulated as

\begin{equation}
\label{multijet}
	\sigma_{\Z/\W}=\sigma_{\Z/\W+0\,\textup{jets}}+\sigma_{\Z/\W+1\,\textup{jets}}+\sigma_{\Z/\W+2\,\textup{jets}}+\sigma_{\Z/\W+n\,\textup{jets}}+...,
\end{equation}

\noindent where $n$ is the number of jets and also sets the lowest power of \as in the expansion of each term. In this study cross-sections for $\sigma_{\Z/\W+0\,\textup{jets}}$ and $\sigma_{\Z/\W+1\,\textup{jets}}$ are computed to NLO, while LO MEs are used for $\sigma_{\Z/\W+2\,\textup{jets}}$. Provided the jets are energetic and well-separated, this serves as a reasonable approximation to the multijet cross-section. However, a PS is needed to extend the jet multiplicity beyond that, to describe collinear and infrared emissions, and to provide a natural entry point into hadronisation models. The basics of the PS formalism will now be outlined.

\section{The parton shower framework}
\label{sec:partonShowerFramework}

While the previous section discusses the issues involved with performing a precise pQCD calculation at fixed order in perturbation theory, this section presents an alternative approach where an approximate calculation is carried out to all orders in \as. A calculation that encompasses higher orders is particularly relevant for collinear branchings and the emission of soft partons. Splittings of the type $q\to qg$, $\bar{q}\to \bar{q}g$, $g\to q\bar{q}$ will be considered, all with QED counterparts, along with the $g\to gg$ branching which stems from the non-Abelian nature of QCD\footnote{While the formalism outlined in this section deals with QCD emissions, it should be emphasised that QED emissions can be handled in a similar manner.}. The tree-level MEs for these branchings are divergent unless virtual corrections are computed, with the exception of $g\to q\bar{q}$ which does not have an infrared divergence (but does still have a collinear divergence).

\begin{figure}[tb]
\begin{center}
	\includegraphics[width=0.9\linewidth]{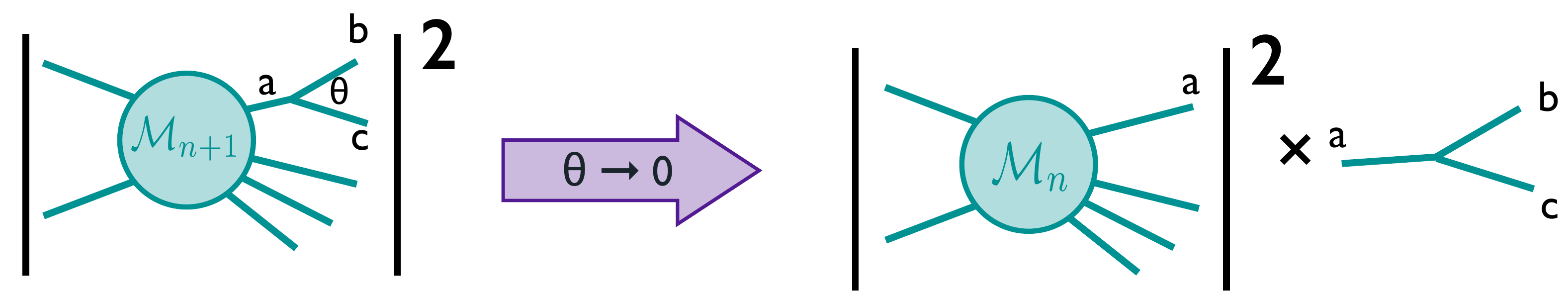}
\vspace*{-0.5cm}
\end{center}
\caption{
	The production of $n+1$ outgoing particles can be described as the product of the $n$-particle ME and the $a \to b c$ branching provided $\theta \to 0$. Figure taken from~\cite{collinearFactorisationFigure}.
	}
\label{fig:collinearFactorisation}
\end{figure}

To begin, consider a hard scattering interaction that leads to $n+1$ outgoing partons. Suppose that within this process particle $a$ branches to $b$ and $c$ at an angle $\theta$. Letting $\theta\to 0$, $a$ becomes nearly on-shell with a time scale for the branching that is considerably longer than the one characteristic of the hard subprocess. In this limit, the interaction can be described as a product of the ME for the production of $n$ particles and the $a\to bc$ branching as illustrated in Fig.~\ref{fig:collinearFactorisation}. Formally, the collinear factorisation of the interaction can be expressed as

\begin{equation}
  \mathopen|\mathcal{M}_{n+1}\mathclose|^2d\Phi_{n+1}\approx  \mathopen|\mathcal{M}_{n}\mathclose|^2\Phi_{n}\frac{dt}{t}dz\frac{d\xi}{2\pi}\frac{\as}{2\pi}P_{a\to bc}(z),
\end{equation}

\noindent where $\mathcal{M}$ is a ME, $\Phi$ denotes phase space, and other terms are explained in what follows.

The branchings are characterised by the energy fraction $z=E_b/E_a$ carried by one of the emerging partons, and an ordering (or evolution) variable $t$ such that each subsequent emission has decreasing $t$. There is some freedom in the choice of the physical form of $t$. One possibility is to use the virtuality of the mother parton while other common choices include transverse momentum and the energy-weighted opening angle of emission. Here, $\xi$ is the angle between the polarisation of $a$ and the plane of the branching. The functions $P_{a\to bc}(z)$ are known as the Dokshitzer-Gribov-Lipatov-Altarelli-Parisi~\cite{DGLAP1,DGLAP2,DGLAP3} splitting kernels. The spin-averaged kernels are

\begin{center}
	\includegraphics[width=0.8\linewidth]{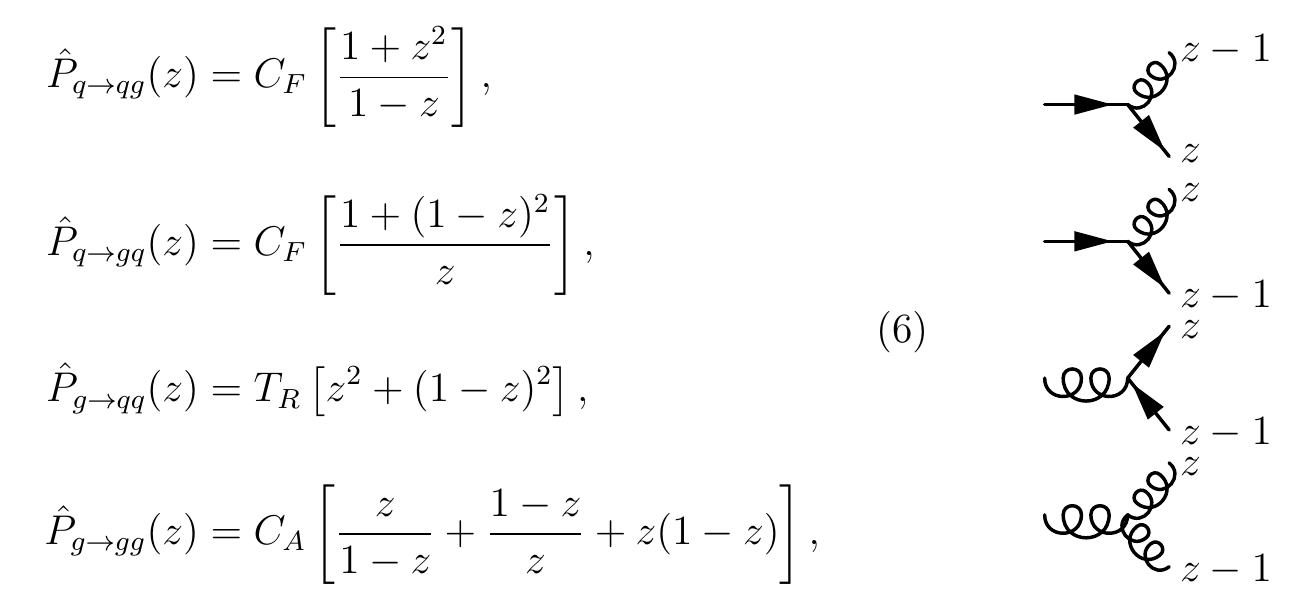}
\end{center}

\noindent where $C_F=\frac{4}{3}$, $C_A=3$, and $T_R=\frac{1}{2}$. Eqs.~6 illustrate that gluons radiate the most, and that the $g\to qq$ splitting does not have a soft divergence.

Proceeding by considering $\mathcal{M}_{n+1}$ as the core process with an additional $b\to de$ branching, the differential \xsec becomes
\setcounter{equation}{6}
\begin{align}
  \begin{split}
    \mathopen|\mathcal{M}_{n+2}\mathclose|^2d\Phi_{n+2}\approx  \mathopen|\mathcal{M}_{n}\mathclose|^2\Phi_{n}&\frac{dt}{t}dz\frac{d\xi}{2\pi}\frac{\as}{2\pi}P_{a\to bc}(z) \\
\times &\frac{dt'}{t'}dz'\frac{d\xi'}{2\pi}\frac{\as}{2\pi}P_{b\to de}(z).
  \end{split}
\end{align}

\noindent This procedure can be extended to an arbitrary number of emissions, each with an independent branching probability.

\begin{figure}[tb]
\begin{center}
	\includegraphics[width=0.6\linewidth]{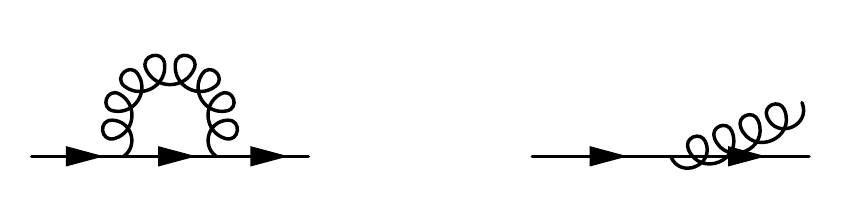}
\vspace*{-0.5cm}
\end{center}
\caption{
	Virtual corrections and unresolvable emissions are accounted for via a Sudakov form factor which gives the probability for no branching to occur between two points along the PS evolution.}
\label{fig:Sudakov}
\end{figure}

Having established the machinery for real emissions, virtual corrections and non-resolvable emissions (as in Fig.~\ref{fig:Sudakov}) must also be accounted for such that unitarity is ensured and divergences are counterbalanced. This is achieved by computing a non-emission probability, known as a Sudakov form factor, that is conventionally labelled as $\Delta$. The differential probability, $\mathcal{P}$, for a branching to not occur at scale $t$ is

\begin{equation}
  d\mathcal{P}_{non-emission}(t)=1-d\mathcal{P}_{emission}(t)=1-\frac{dt}{t}\frac{\as}{2\pi}\int{dz\,\hat{P}(z)}.
\end{equation}

\noindent Therefore, the probability for no branchings to occur between scales $t_1$ and $t_2$ split into $N$ infinitesimal subranges, \ie the Sudakov form factor, is

\begin{align}
  \begin{split}
    \Delta (t_1,t_2)&=\lim_{N \to \infty}{\prod_{i=0}^N{\left(1-\frac{dt}{t_i}\frac{\as}{2\pi}\int{dz\,\hat{P}(z)}\right)}}, \\
    &\simeq \lim_{N \to \infty} \exp{\left(\sum_{i=0}^N{\left(-\frac{dt}{t_i}\frac{\as}{2\pi}\int{dz\,\hat{P}(z)}\right)}\right)}, \\
    &\simeq \exp{\left(-\frac{\as}{2\pi}\int_{t_1}^{t_2}{\frac{dt}{t}}\int{dz\,\hat{P}(z)}\right)}.
  \end{split}
\end{align}

With the emission and non-emission probabilities specified, the cascade of partons can be evolved down to a certain virtuality, conventionally $\mathcal{O}(1\gev)$, and a hadronisation model is then applied. A MC event generator implements this model for final-state showers as a Markov chain \ie with independent randomly generated splittings. Initial-state showers are treated in a similar fashion working backwards from the hard interaction to earlier branchings with the added complication that a PDF weight must be applied to each branching.

The PS framework thus offers a robust, physical way of treating multiple splittings to an arbitrary number of partons. The formalism can be applied to generate a timelike shower to describe jets in the final state or evolved backward from the hard interaction to model initial-state radiation. Crucially, it can be shown that the effect of a PS is unitary \ie it leaves the inclusive \xsec unchanged~\cite{UMEPS}. The PS also gives an approximate estimate of the effect of higher order corrections that can become large at edges of phase space. In addition, it is a natural gateway to the non-perturbative regime of QCD where hadronisation models are employed. Finally, the method is well suited for computer implementation and forms the basis for several MC event generators.

As the PS approximation is strictly true only in the collinear limit and does not fully account for quantum interference, input from a precise ME calculation is typically needed to improve the description of well-separated jets and measures of event shape. This marriage of a ME calculation to a PS is the focus of the next section.

\section{Matching matrix elements to parton showers}
\label{sec:matchingMerging}

\begin{table}[t]
\renewcommand{\arraystretch}{2}
\setlength{\tabcolsep}{25pt}
\begin{center}
\begin{tabular}{ll}
    \hline
    Matrix element                           												& Parton shower       \\ 
    \hline
    \textcolor{ForestGreen}{\pbox{20cm}{Good description of \\ well-separated hard partons}} 	& \textcolor{ForestGreen}{Resums to all orders} \\
    \textcolor{ForestGreen}{\pbox{20cm}{Exact to given order \\ in perturbation theory}}		& \textcolor{ForestGreen}{\pbox{20cm}{Arbitrary particle multiplicity $\Rightarrow$ \\ ideal for description of jet formation}} \\
    \textcolor{ForestGreen}{Quantum interference correct}                                                         	& \textcolor{ForestGreen}{Hadronisation model easily applied} \\
    \textcolor{Red}{Limited number of particles}             								& \textcolor{ForestGreen}{Computationally cheap} \\
    \textcolor{Red}{Computationally expensive} 										& \textcolor{Red}{Valid in collinear limit} \\
    \textcolor{Red}{\pbox{20cm}{Hadronisation model difficult \\ to implement}} 									& \textcolor{Red}{\pbox{20cm}{Quantum interference \\ not fully accounted for}} \\
    \hline
  \end{tabular}
\end{center}
\caption{
    Pros and cons of ME and PS descriptions of jet formation.}
\label{tab:MMProsCons}
\end{table}

As the previous sections illuminate, ME and PS calculations have different virtues and ranges of applicability. The strengths and weaknesses of both approaches are summarised in Table~\ref{tab:MMProsCons}. Ultimately, it is desirable to accurately describe both the hard process and the development of partons into jets. This achieved by interfacing a ME calculation and a PS. The combination would use a $\Z/\W+(0,1,...,n)$ parton ME for events with up to $n$ hard jets and dress up the outgoing partons with radiation using the PS. Given that $n$ is finite, additional jets must then arise from PS emissions. In this manner, a more accurate description of events with up to $n$ jets is achieved and the precision of the inclusive \xsec is improved.

Specifically, a given $(m+1)$-jet event, where $0 \leq m < n$, can be obtained in two ways: from the PS evolution of a $(m+1)$-parton ME without branchings hard enough to produce a jet; or from a $m$-parton ME, where the PS leads to the formation of an additional jet. A particular \mnm scheme defines which of the two paths should be followed on an event-by-event basis.

\begin{figure}[tb]
\begin{center}
	\includegraphics[width=0.45\linewidth]{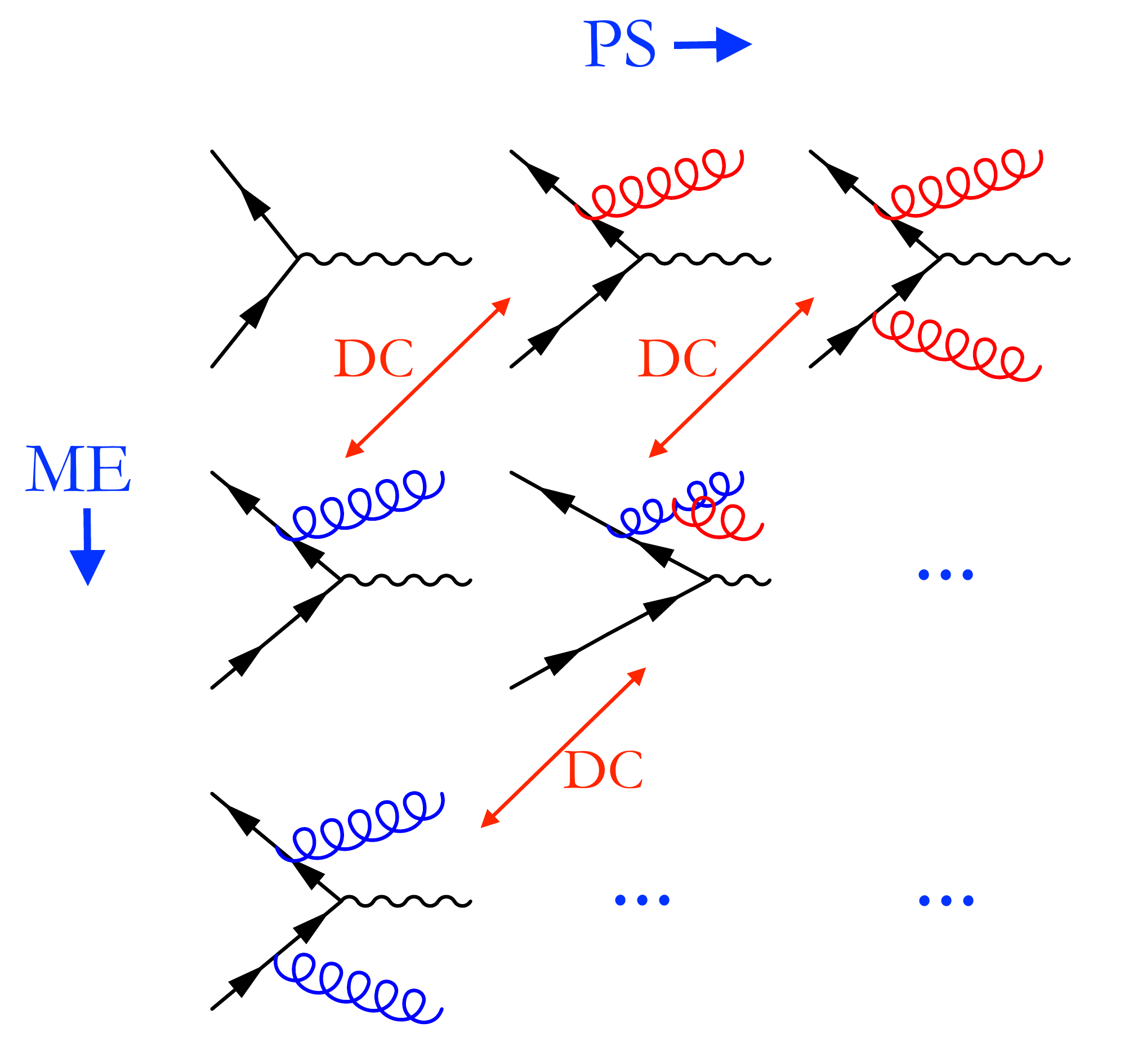}
\vspace*{-0.5cm}
\end{center}
\caption{
	Branchings in the PS (in red) and real emissions in a ME calculation (in blue) can lead one to double count (DC) configurations with the same number of outgoing particles.}
\label{fig:DCFig}
\end{figure}

The potential pitfalls with this approach are double counting and the biased filling of phase space. Double counting arises when an identical final state is allowed to be generated by both a $(m+1)$-parton ME and a $m$-parton ME with an extra jet from the PS. This is illustrated for increasing multiplicities in Fig.~\ref{fig:DCFig}. The solution is to introduce a phase-space separation, referred to as a matching scale, that vetoes PS emissions above a certain threshold. In other words, the PS is only allowed to generate unresolved emissions for multiplicities for which there is a ME calculation available. 

In this study, the ME-based sample was required to have jets with transverse momentum, \jetpt, in excess of 10\gev, where the jets are clustered using the longitudinally-invariant \kt-clustering algorithm for hadron-hadron collisions~\cite{kTalgorithm}. The \kt-scheme defines two final-state particles $i$ and $j$ as belonging to separate jets if their relative transverse momentum squared,

\begin{equation}
	k^2_{\text{T}ij}=2\min{\left(\pt_i,\pt_j\right)^2}\,\frac{\left(\cosh(\eta_i-\eta_j)-\cos(\phi_i-\phi_j)\right)}{R^2},
\end{equation}

\noindent is larger than $\left(10 \gev\right)^2$, where $R$ is a parameter that controls the size of the jet, $\eta$ is the pseudorapidity and $\phi$ is the azimuthal angle. The threshold effectively controls the relative weights given to the ME calculation and the PS. A low threshold enhances the contribution from the ME while a high scale gives a bigger role to the shower approximation.

\begin{figure}[tb]
\begin{center}
	\includegraphics[width=0.38\linewidth]{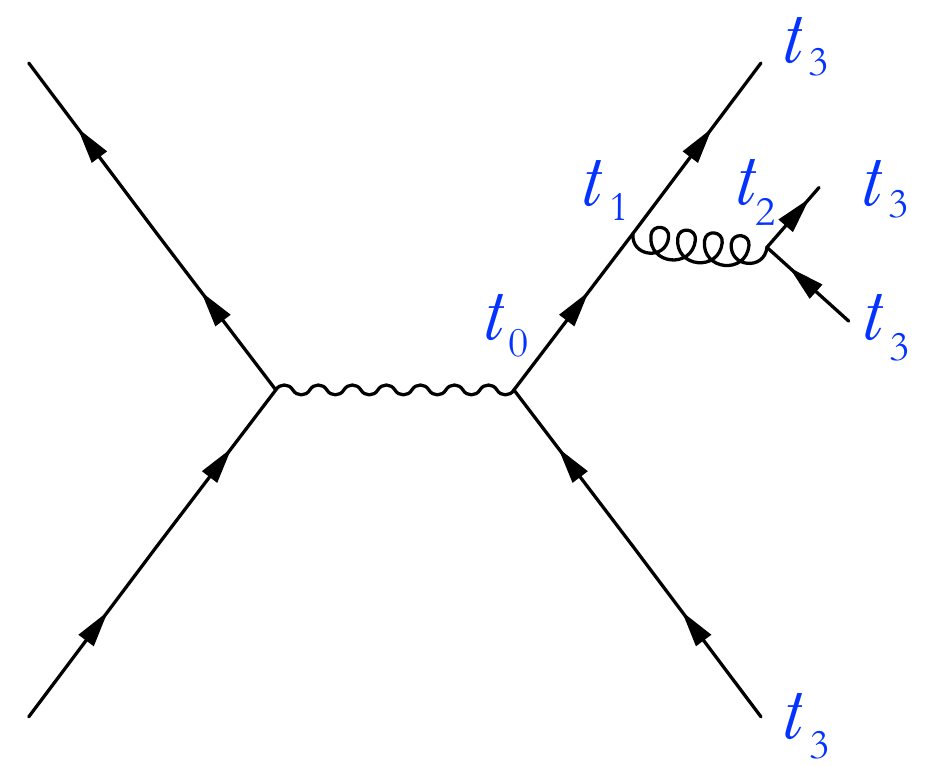}
\vspace*{-0.5cm}
\end{center}
\caption{
	A $2\to 4$ process with gluon emission and subsequent splitting to illustrate the relevant scales for \as reweighting and Sudakov suppression of the ME.}
\label{fig:MEcorrection}
\end{figure}

One must also make sure that there is a smooth transition along the phase space boundary between the ME calculation and the PS. This is achieved by two modifications to the ME. Firstly, the ME must be reweighted for the value of \as that would have been used had the configuration been generated by a PS, which uses a running value of \as. Secondly, the PS employs Sudakov factors for the non-emission probability. This requires a pseudo-PS history to be created for each ME-based event and a Sudakov reweighting applied to it. This history interprets a particular final state as the result of a sequence of branchings in a PS. Clearly, there is no unique history for a given event so all histories must be generated and one chosen probabilistically. For the configuration shown in Fig.~\ref{fig:MEcorrection} with a starting scale of $t_0$ and emissions at $t_1$ and $t_2$, the ME becomes

\begin{equation}
	\mathopen|\mathcal{M}\mathclose|^2 \to \mathopen|\mathcal{M}\mathclose|^2\frac{\as(t_1)}{\as(t_0)}\frac{\as(t_2)}{\as(t_0)}
\left(\Delta_q(t_{3},t_0)\right)^2\Delta_g(t_2,t_1)\left(\Delta_q(t_{3},t_2)\right)^2,
\end{equation}

\noindent where $t_{3}$ is the endpoint of the evolution in this example. Sudakov suppression also serves to make MEs exclusive such that different parton multiplicities can be added \ie merged.

Several prescriptions have been proposed to resolve the issues of avoiding double counting and ensuring a smooth transition between hard and soft regimes. Additionally, these approaches differ in their treatment of higher orders in the ME calculation. One strategy is to extend real emissions to higher orders at LO accuracy while approximating virtual/loop corrections with shower Sudakov factors. Alternatively, NLO MEs can be used for the first few emissions. The \ckkw~\cite{CKKW}, \ckkwl~\cite{CKKWL1,CKKWL2,CKKWL3,CKKWL4}, \mlm~\cite{MLM}, and \umeps~\cite{UMEPS} schemes employ LO MEs while NLO MEs are used in \fxfx~\cite{FXFX} and \unlops~\cite{UNLOPS}. Finally, NNLO MEs have been employed for Drell-Yan and Higgs physics in the \unnlops scheme~\cite{UNNLOPS1,UNNLOPS2}.

What follows is an overview of the different \mnm schemes that were included in this study with emphasis on the differences between them.

\subsection{The \mlm scheme}

Two earlier \mnm schemes, \ckkw and \ckkwl, examine individual branchings within a PS and reject those with $t_i > t_{\textup{MS}}$ for branching $i$ and matching scale $t_{\textup{MS}}$ to establish a phase space division. However, this form of truncated showering requires a modification of the showering algorithm. The innovation of the MLM scheme is to evolve the PS from $t_0$ to the hadronisation scale, form jets and reject the event if a shower-initiated jet exceeds the matching scale. The scheme is attractively simple by enabling matching to any shower algorithm without requiring modifications. However, the Sudakov suppression of the ME is, as a consequence, imprecise since a product of Sudakov factors that the jet represents is substituted for the factor associated to a single branching.

\subsection{Unitarised matrix element + parton shower merging}

As previously mentioned, it can be shown that the PS does not modify the total cross-section - a property known as PS unitarity. However, since there is a only a partial correspondence between $m$-jet MEs and an equivalent sequence of approximate splitting kernels, the total cross-section is modified in the \ckkwl \mnm scheme. Specifically, the PS is generally correct to LL or NLL accuracy. Therefore, a residual mismatch remains for further subleading terms.

\umeps resolves this mismatch by explicitly integrating over the phase space of a jet in the $m$-jet sample to arrive at a virtual correction to the $(m-1)$-jet sample. This enables the scheme to better describe unresolvable emissions at any multiplicity and enforces a cancellation between real and virtual terms for any multiplicity. In this manner the unitarity of the inclusive cross section is enforced. The cancellation of unresolved and resolved contributions also significantly reduces the scheme's dependence on the choice of matching scale.

\subsection{The \fxfx scheme}

If all jets in a process are the result of PS emissions, one obtains an inclusive cross-section which is LO+LL accurate with LL accurate exclusive observables\footnote{If the definition of a given observable explicitly involves $m$ jets, it is considered exclusive in $m$ jets and inclusive in $n-m$ jets with $0 \leq m \leq n$. When $m=0$, the observable is typically called fully inclusive.}. The purpose of tree-level \mnm schemes is to improve the description of exclusive observables with LO MEs. The aim of the \fxfx scheme is to promote exclusive observables to NLO+LL accuracy. This is advantageous as the precision of collider phenomenology is improved and scale uncertainties are greatly reduced. In the \fxfx scheme, the $m$-parton cross-section is calculated using the \mcatnlo procedure~\cite{MCatNLO} and the matching to the PS is handled according to the MLM procedure. Unlike other schemes, the matching scale can be replaced by a smooth monotonic function (effectively, a matching range).

\subsection{Merging NLO matrix elements with parton showers}

Multi-jet tree-level merging improves event shapes but cannot give a correct overall normalisation or decrease the uncertainty deriving from scale variations. The \unlops method aims to have a NLO accurate description of exclusive $m$-jet final states while also ensuring the unitarity of the overall cross-section in a manner inspired by the \umeps scheme. In this spirit, once a one-jet NLO calculation is added, its integrated version is used to correct zero-jet events and similarly for all multiplicities that MEs are available for. The precision of the method is further improved by including tree-level MEs for multiplicities beyond those for which an NLO calculation is available. These multiplicities are handled in a manner similar to that of \umeps. Analogously to \umeps, variations caused by a dependence on the matching scale are minimised due to the ``subtract what you add" principle. Finally, it is also worth pointing out that the scheme can be straightforwardly promoted to NNLO accuracy (termed \unnlops) as is established in Refs.~\cite{UNNLOPS1,UNNLOPS2}.

\subsection{The appropriate choice of PDF}

A NLO ME calculation gives an enhancement of a factor $\ln{\left(1/x\right)}$ that is compensated by a commensurate term in a NLO PDF. It is crucial that the order of the ME and the PDF match, particularly for small $x$ where $\ln{\left(1/x\right)}$ is large. Therefore, a LO PDF is used for the PS and MPI.

\section{Results on matching and merging}
\label{sec:resultsMatchingMerging}

This study is carried out using \pythia 8.201~\cite{Pythia82}. Events based on MEs are generated externally using the \amcatnlo framework as implemented in \madgraph 5.2.3.2.2~\cite{madgraph} and interfaced to \pythia using LHE-files~\cite{LHE}. The NLO \mnm schemes have a NLO CT10 PDF set~\cite{CT} in the ME-level calculation, while the LO CTEQ6L1 PDF set~\cite{CTEQ} is employed for LO \mnm schemes. The CTEQ6L1 PDF set is also used for the PS and for MPI within \pythia. \as is set to 0.118 at the \Z mass in the ME calculation and a value of 0.1365 is used for modelling the remainder of the event. The matching scale is set at 10\gev. All comparisons to LHC measurements are carried out within the \rivet framework~\cite{Rivet}.

The standard LO description of weak gauge boson production is referred to as the ``Born description" in what follows. \mlm is LO up to 2 jets; \fxfx is NLO to 1 jets; \unlops is NLO to 1 jets and LO up to 2 jets. Higher jet multiplicities are generated by the PS. In the central region the predictions are compared to measurements by ATLAS~\cite{ATLASZjets} and CMS~\cite{CMSZeventShapes} at $\sqrt{s}=7$\tev. At the same centre-of-mass energy, a \lhcb measurement of $\Z+$jets~\cite{LHCbZjets} is used to gauge agreement in the forward region. Finally, the predictions are compared to a measurement of inclusive \Z production in the forward region at $\sqrt{s}=13$\tev~\cite{Z13TeV}.

\begin{figure}[tb]
\begin{center}
\begin{minipage}{.5\textwidth}
	\centering
	\subfigure[]{
		\includegraphics[width=0.95\linewidth]{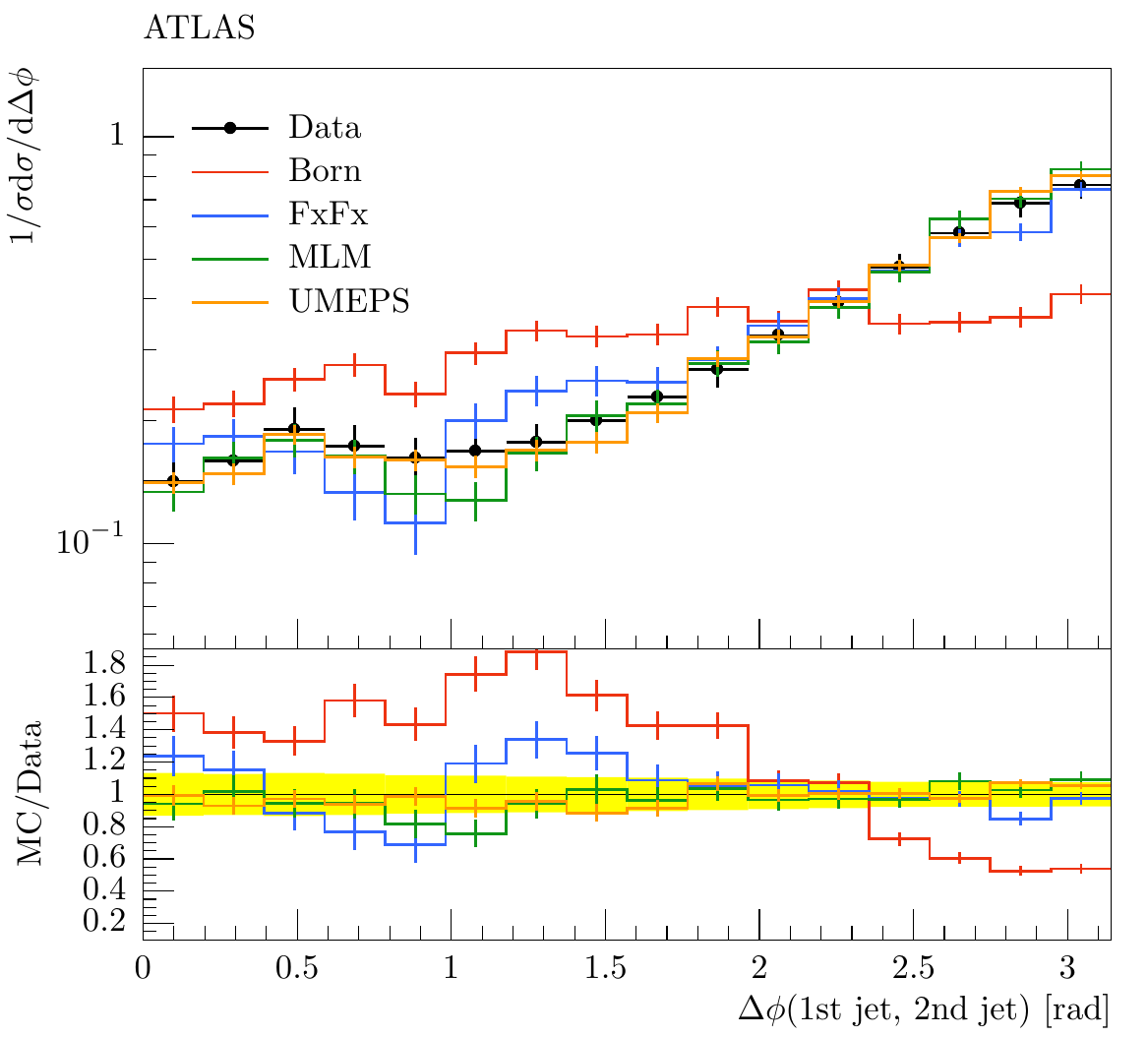}
		\vspace*{-0.5cm}
	}
\end{minipage}
\hspace{-0.25cm}
\begin{minipage}{0.5\textwidth}
	\centering
	\subfigure[]{
		\includegraphics[width=0.95\linewidth]{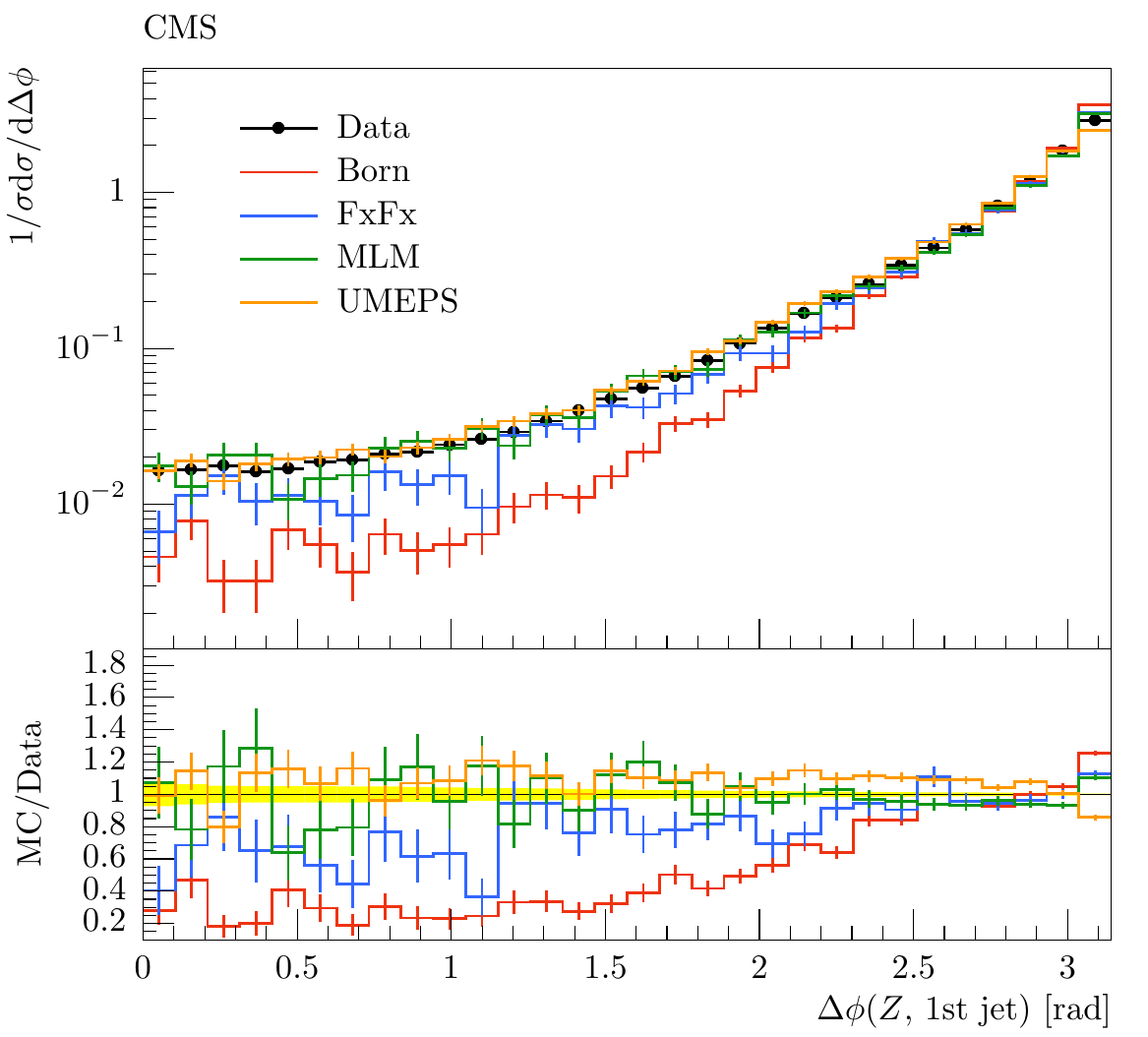}	
		\vspace*{-0.5cm}
	}
\end{minipage}
\end{center}
\caption{The normalised differential \xsec for $\Z+$jets production in leptonic final states as a function of the (a) azimuthal angle between the leading and subleading jet and (b) the \Z boson and the leading jet as measured by (a) ATLAS~\cite{ATLASZjets} and (b) CMS~\cite{CMSZeventShapes}. Data points with full uncertainties are compared to the Born description in red and a selection of \mnm schemes. Simulation uncertainties are statistical only.}
\label{fig:centralEventShapes}
\end{figure}

Fig.~\ref{fig:centralEventShapes} illustrates that \mnm schemes clearly outperform the Born description in modelling event shapes such as the angle between jets or between the \Z boson and the jets. No significant differences between the \mnm schemes are observed.

\begin{figure}[tb]
\begin{center}
\begin{minipage}{.5\textwidth}
	\centering
	\subfigure[]{
		\includegraphics[width=0.95\linewidth]{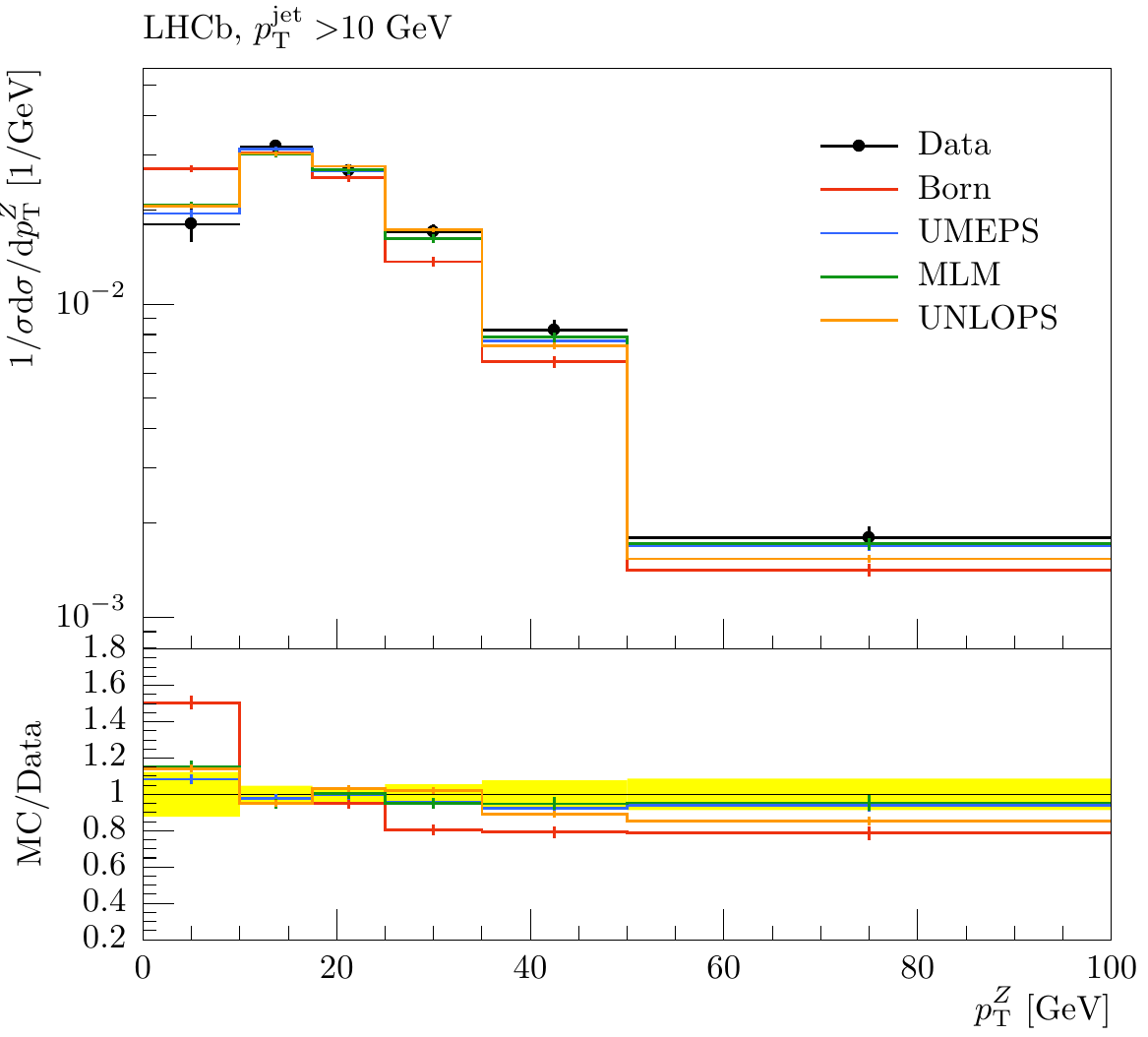}
		\vspace*{-0.5cm}
	}
\end{minipage}
\hspace{-0.25cm}
\begin{minipage}{0.5\textwidth}
	\centering
	\subfigure[]{
		\includegraphics[width=0.95\linewidth]{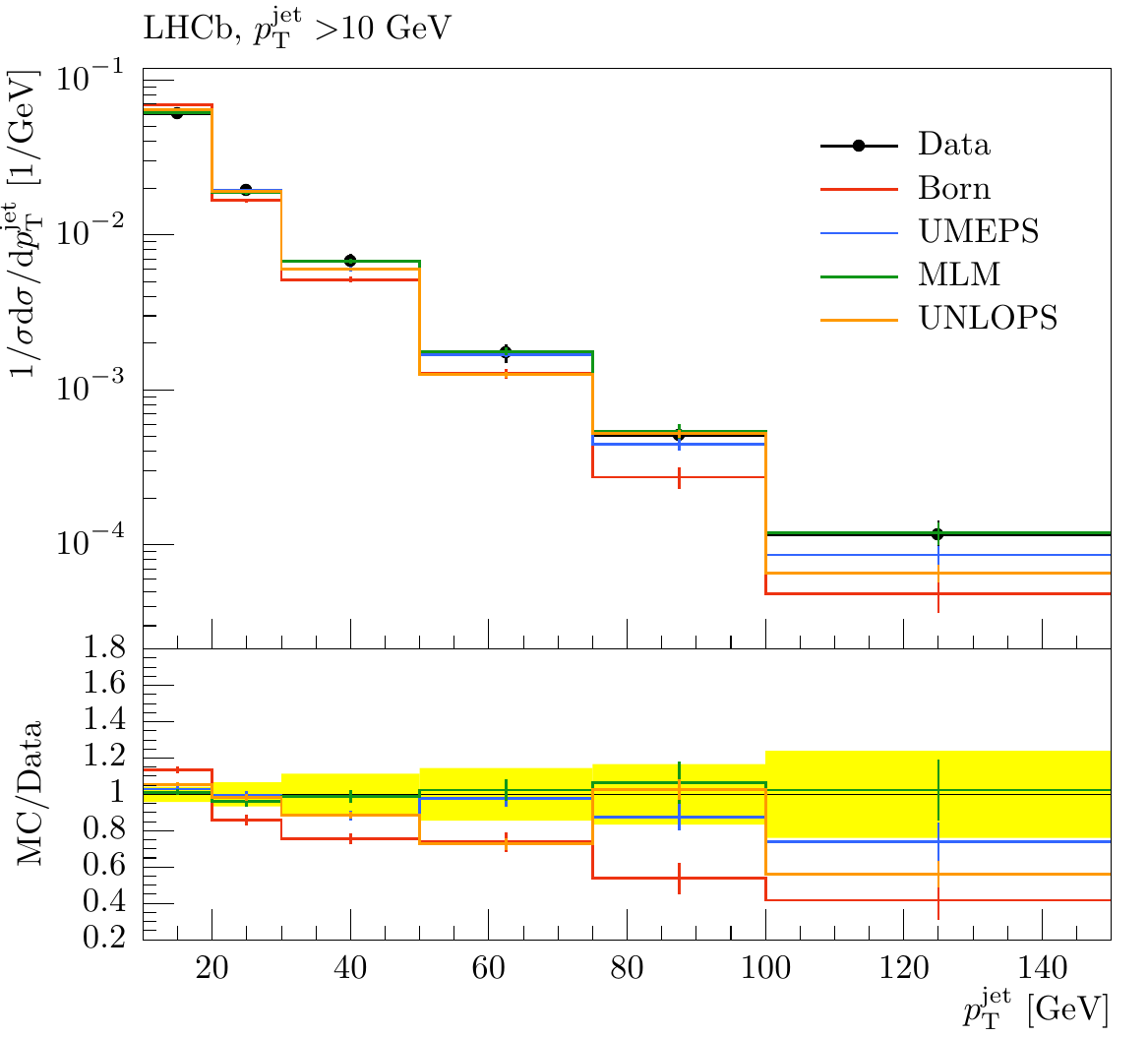}	
		\vspace*{-0.5cm}
	}
\end{minipage}
\end{center}
\caption{The normalised differential \xsec for $\Z+$jets production in the muon final state as a function of the (a) transverse momentum of the \Z and (b) the transverse momentum of the leading jet as measured by \lhcb~\cite{LHCbZjets}. Data points with full uncertainties are compared to the Born description in red and a selection of \mnm schemes. Simulation uncertainties are statistical only.}
\label{fig:forwardZjetpT}
\end{figure}

\begin{figure}[tb]
\begin{center}
\begin{minipage}{.5\textwidth}
	\centering
	\subfigure[]{
		\includegraphics[width=0.95\linewidth]{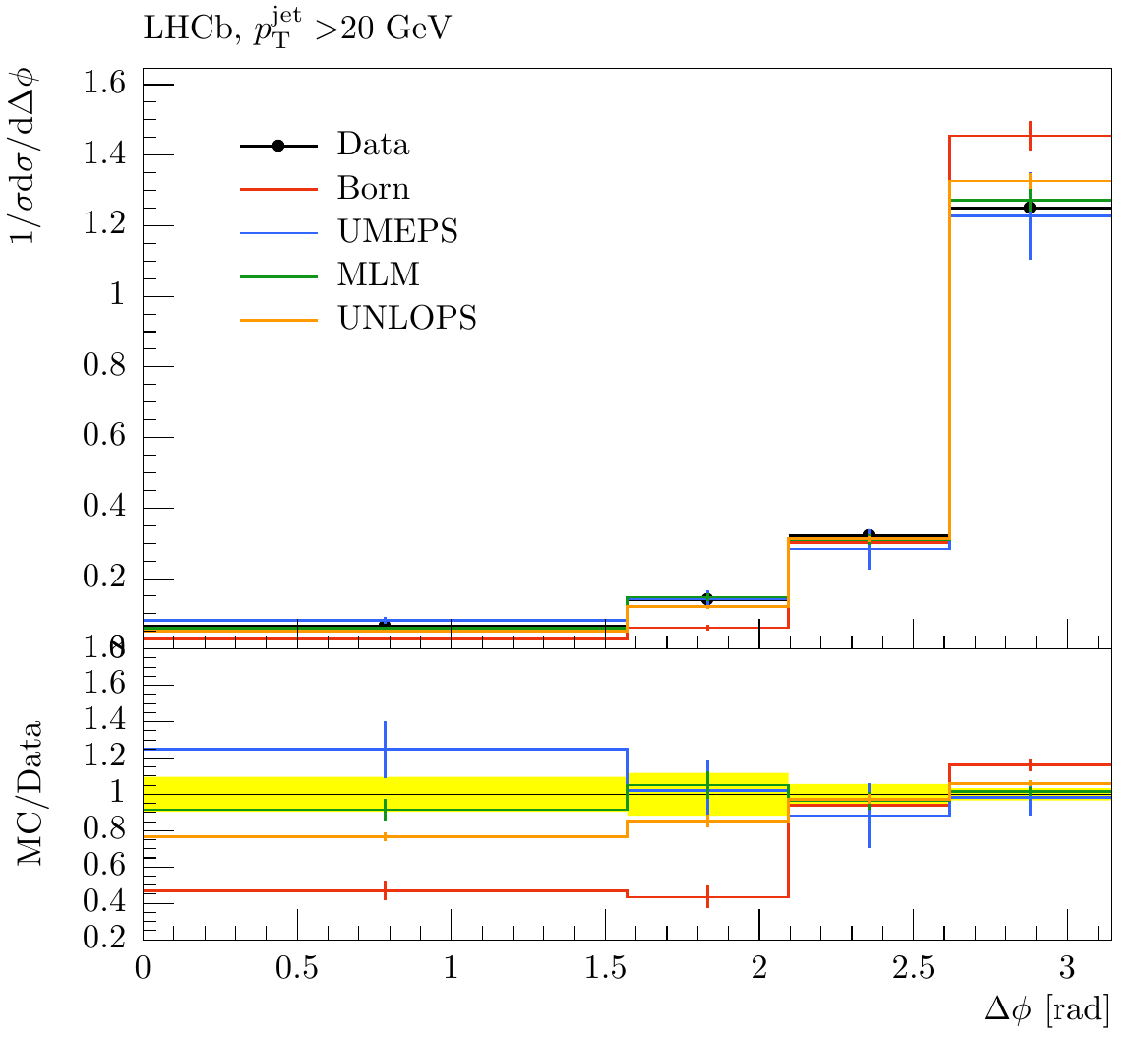}
		\vspace*{-0.5cm}
	}
\end{minipage}
\hspace{-0.25cm}
\begin{minipage}{0.5\textwidth}
	\centering
	\subfigure[]{
		\includegraphics[width=0.95\linewidth]{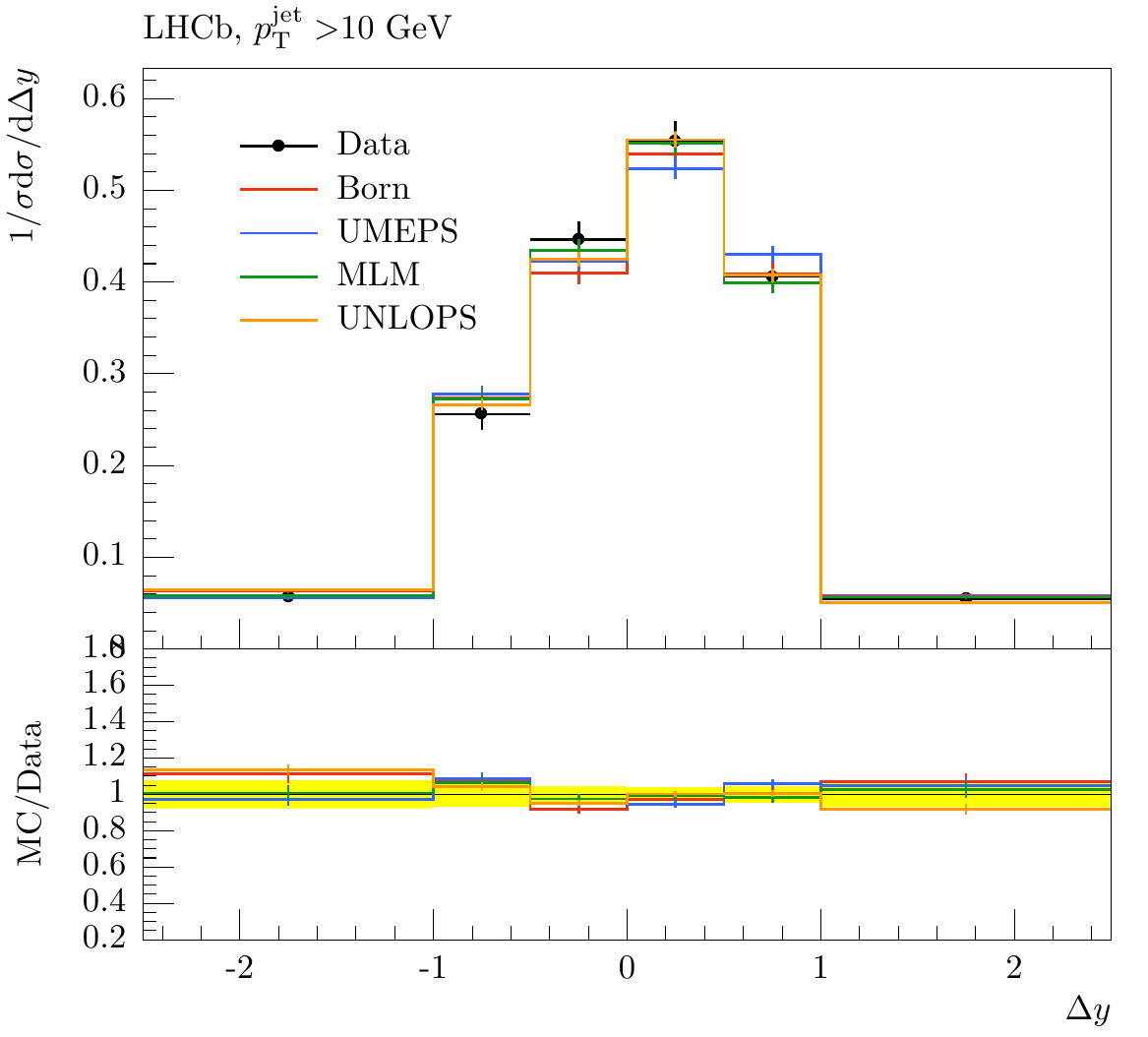}	
		\vspace*{-0.5cm}
	}
\end{minipage}
\end{center}
\caption{The normalised differential \xsec for $\Z+$jets production in the muon final state as a function of the (a) azimuthal angle and (b) the rapidity between the \Z boson and the leading jet as measured by \lhcb~\cite{LHCbZjets}. Data points with full uncertainties are compared to the Born description in red and a selection of \mnm schemes. Simulation uncertainties are statistical only.}
\label{fig:forwardZjeteventShapes}
\end{figure}

Figs.~\ref{fig:forwardZjetpT} and~\ref{fig:forwardZjeteventShapes} (a) illustrate that certain regions of phase space are poorly modelled in the forward region by the Born description. Discrepancies are seen in the first bin of Fig.~\ref{fig:forwardZjetpT} (a), which corresponds to events with multiple hard jets, but low \Z boson \pt, \ptZ. Also, the slope of the differential \xsec as a function of \jetpt is incorrect. Due to momentum conservation, mismodelling of \jetpt also has an impact on \ptZ and, by extension, the \pt of the daughter leptons. Finally, the Born description predicts a larger number of pencil-like event topologies when compared to the measurement and to the predictions from \mnm schemes as seen in Fig.~\ref{fig:forwardZjeteventShapes} (a). Fig.~\ref{fig:forwardZjeteventShapes} (b) gives the difference in rapidity between the \Z boson and the leading jet with no significant differences seen between the predictions. Overall, it is clear that \mnm schemes offer a superior description of event topologies with multiple hard jets in the forward region.

\begin{figure}[tb]
\begin{center}
\begin{minipage}{.5\textwidth}
	\centering
	\subfigure[]{
		\includegraphics[width=0.95\linewidth]{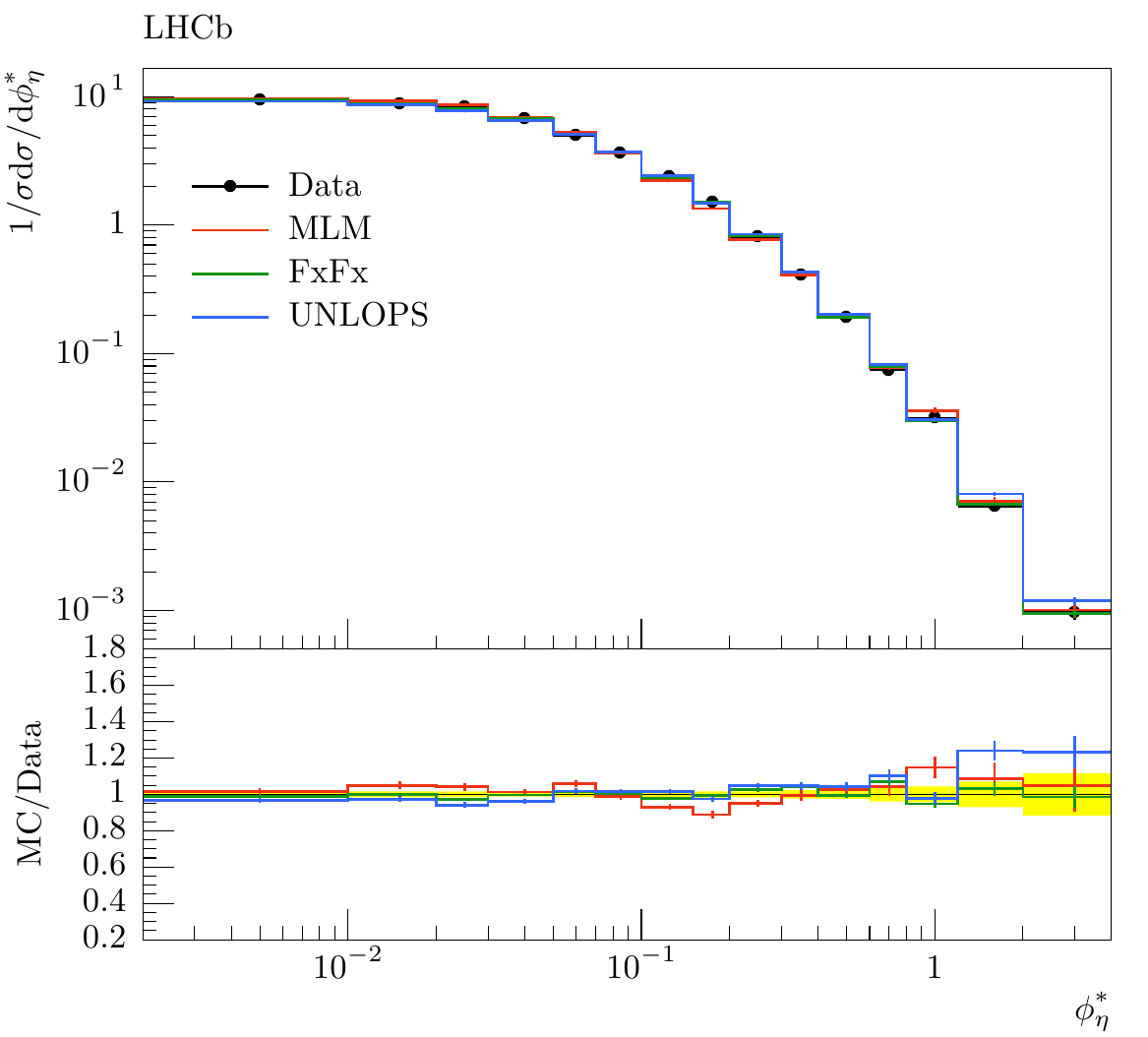}
		\vspace*{-0.5cm}
	}
\end{minipage}
\hspace{-0.25cm}
\begin{minipage}{0.5\textwidth}
	\centering
	\subfigure[]{
		\includegraphics[width=0.95\linewidth]{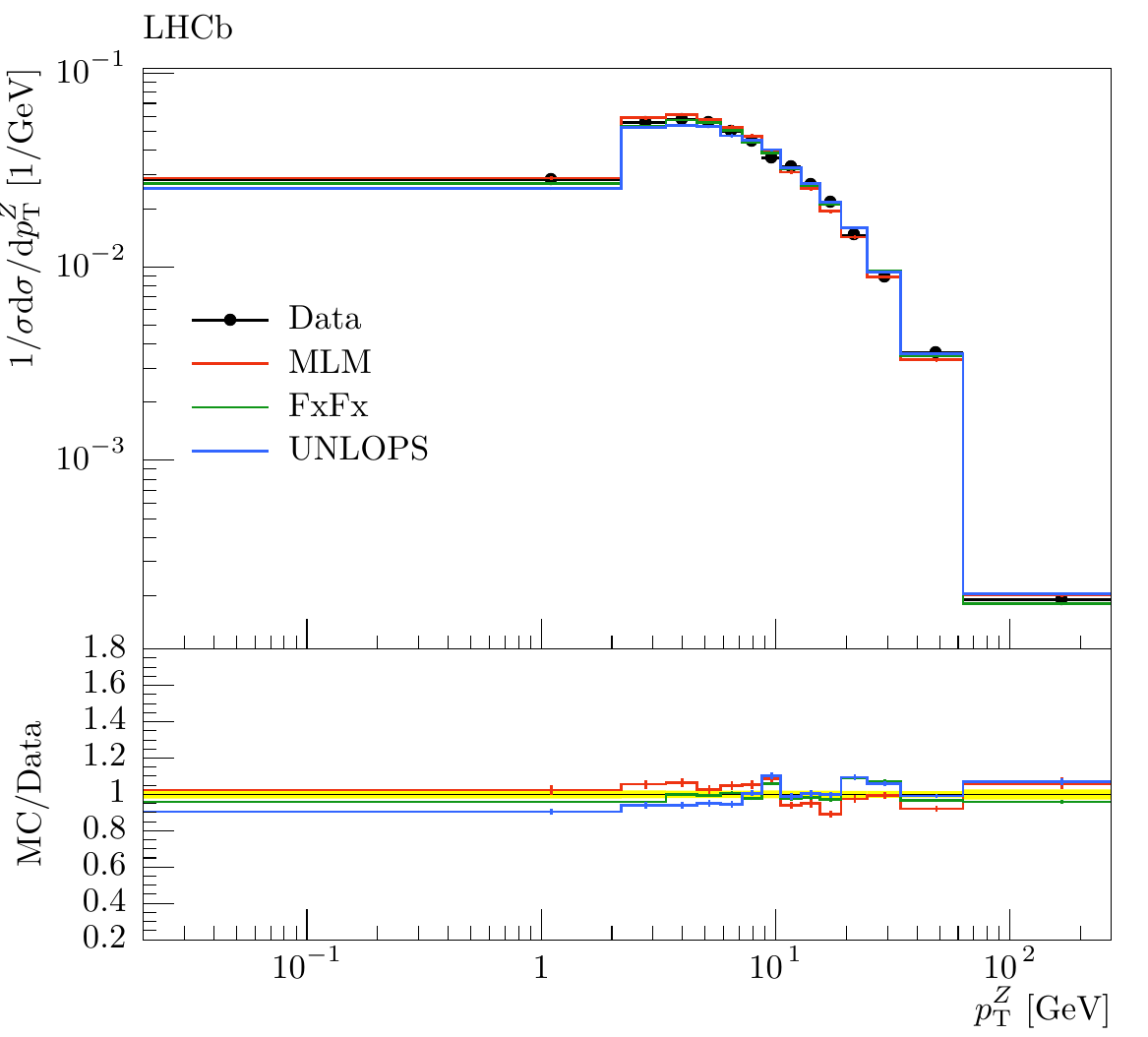}	
		\vspace*{-0.5cm}
	}
\end{minipage}
\end{center}
\caption{The normalised differential \xsec for inclusive \Z boson production in the muon final state as a function of (a) $\phi^{*}_{\eta}$ and (b) the transverse momentum of the \Z as measured by \lhcb~\cite{Z13TeV}. Data points with full uncertainties are compared to a selection of \mnm schemes. Simulation uncertainties are statistical only.}
\label{fig:forwardincZ}
\end{figure}

Fig.~\ref{fig:forwardincZ} illustrates that \mnm schemes model the $\phi^{*}_{\eta}$ distribution\footnote{Relying solely on angles, $\phi^{*}_{\eta}$ acts as a proxy for transverse momentum. It is defined as \mbox{$\phi^{*}_{\eta}\equiv\tan{\frac{\pi-|\Delta\phi|}{2}}/\cosh{\frac{\Delta\eta}{2}}\simeq\pt/M$}, where $M$ and \pt refer to the lepton pair, $\Delta\eta$ and $\Delta\phi$ are the differences in pseudorapidity and azimuthal angles respectively between the leptons.} and \ptZ well. \unlops is slightly discrepant due to the lack of a dedicated tune for NLO \mnm methods in \pythia. Finally, it is worth noting that none of the \mnm schemes have the correct overall \xsec in the forward region with discrepancies at the 5\% level. This suggests corrections beyond NLO are crucial and \unnlops should ideally be used.

\clearpage

\section{Weak gauge boson emission in parton showers}
\label{sec:weakGaugeBoson}

WS approaches \Z and \W production from a novel viewpoint. In a Born description of weak gauge boson production the hardest scale in the event is associated with creating the bosons and any jet activity is modelled by means of a PS originating from initial- or final-state radiation, and MPI. This, however, will typically underestimate the number, and hardness, of associated jets. WS instead starts from a QCD process, most commonly dijet production, and allows for the emission of \Z and \W bosons within the development of the PS~\cite{WeakShowers}. For this purpose, appropriate splitting kernels must be implemented and the helicity of the weak boson daughters accounted for. An additional complication is the relatively high mass of EW bosons which leads to kinematic constraints. In \pythia, energy-momentum conservation is achieved by treating MPI, initial- and final-state radiation together such that all features of the event are in direct competition for phase space~\cite{PhaseSpaceCompetition}.

The Born description and WS can be added to populate the phase space more fully and to achieve a better description of EW physics in an hadronic environment with centre-of-mass energies that can greatly exceed the EW scale. While these two production paths are expected to preferentially populate different regions of phase space, it is clear that their combination can result in double counting. Since both production modes can lead to topologies with the same number of final state jets, care must be taken to veto events which are better described by the competing production path. This is achieved by employing the \kt-clustering algorithm on all final-state particles, including weak bosons. Events with bosons lying close to jets in a Born description and, correspondingly, events with bosons that are separated from jets in the QCD-initiated path are vetoed.

Extending the PS machinery to include EW boson emissions brings several improvements to the simulation of high-energy $pp$ collisions. Experimentally, weak bosons are seen to be present in jets from their leptonic signatures and the emission probability increases as a function of the centre-of-mass energy of the collision. Weak emission, therefore, can lead to a more realistic description of jet substructure which complements the many ongoing experimental efforts to study the internal structure of jets and formulate discriminating variables based on this~\cite{jetSubstructure}.

Secondly, weak corrections are required to reproduce the jet rate suppression expected at \jetpt values in excess of 1\tev. What is more, the WS machinery complements efforts in developing \mnm schemes. As explained previously, \mnm schemes require PS histories to be generated for Sudakov suppression terms. Allowing for EW emission in the PS leads to more realistic histories and also enables histories to be generated for processes which currently lack one.

Finally, this study seeks to test if a better description of EW boson production in association with jets can be achieved with WS. In particular, WS is expected to be relevant for events with hard jets and high jet multiplicities.

\section{Results on weak showering}
\label{sec:resultsWeakShowering}

In this section the Born description of \Z or \W production and simulation of dijet production along with weak emissions within the PS are compared to measurements made at ATLAS~\cite{ATLASZjets}, CMS~\cite{CMSWjets}, and LHCb~\cite{LHCbZjets} at $\sqrt{s}=7$\tev. PDFs are parameterised by the CTEQ6L1 PDF set. The Monash tune~\cite{MonashTune} is used which sets $\as=0.1365$ within the PS. The \Z mass is required to be above $40 \gev$ and PS emissions are allowed up to the kinematic limit of $\pt=\sqrt{s}/2$ to avoid gaps in phase space (known as ``power showers").

The Born description does not yield the correct normalisation of the \xsec and, consequently, an overall scaling is applied (the so-called ``K-factor"). K-factors are chosen empirically by scaling the Born contribution to agree with the data. Centrally, the K-factor for $Z+$jets is taken to be 1.37 based on the zero-jet bin of the exclusive jet multiplicity distribution, \njet. The K-factor for $W+$jets is determined to be 1.28 based on the one-jet bin of the \njet distribution. In the forward region, the Born contribution is scaled by 1.10 to enforce agreement with the $[10-20]\gev$ bin of the $Z+$jet \xsec as a function of \jetpt. Empirically, it is seen that a LO prediction of QCD dijet production does not need a K-factor but rather an enhancement of \as can compensate for missing higher orders~\cite{QCDKFactor}. Here, \ptWZ is required to be above 1\gev with respect to the parton in the shower to remove divergences. Double counting between the Born and WS contributions are removed as described in Sect.~\ref{sec:weakGaugeBoson} with a \kt algorithm $R$ parameter of 0.6.

\begin{figure}[tb]
\begin{center}
\begin{minipage}{.5\textwidth}
	\centering
	\subfigure[]{
		\includegraphics[width=0.95\linewidth]{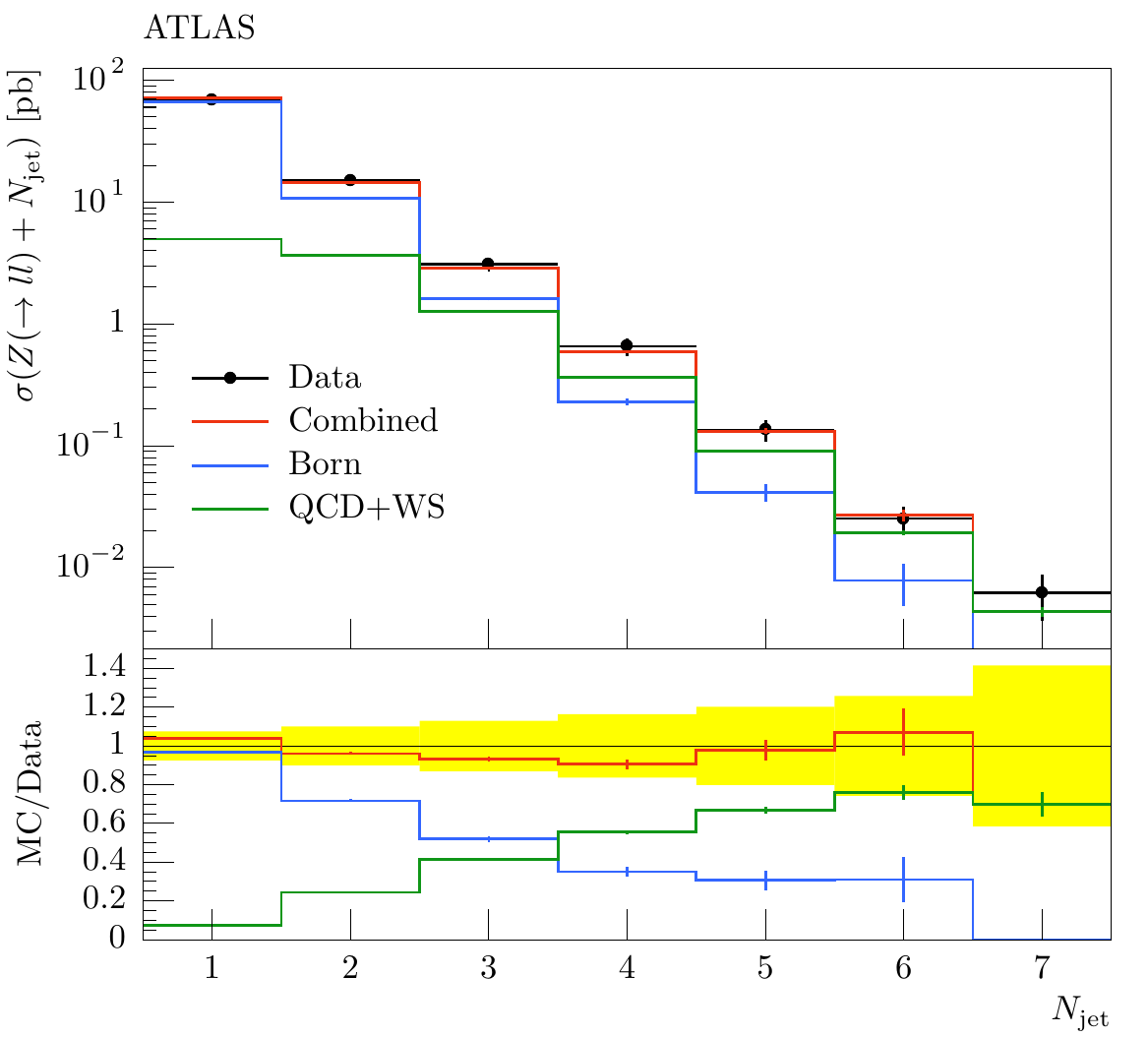}
		\vspace*{-0.5cm}
	}
\end{minipage}
\hspace{-0.25cm}
\begin{minipage}{0.5\textwidth}
	\centering
	\subfigure[]{
		\includegraphics[width=0.95\linewidth]{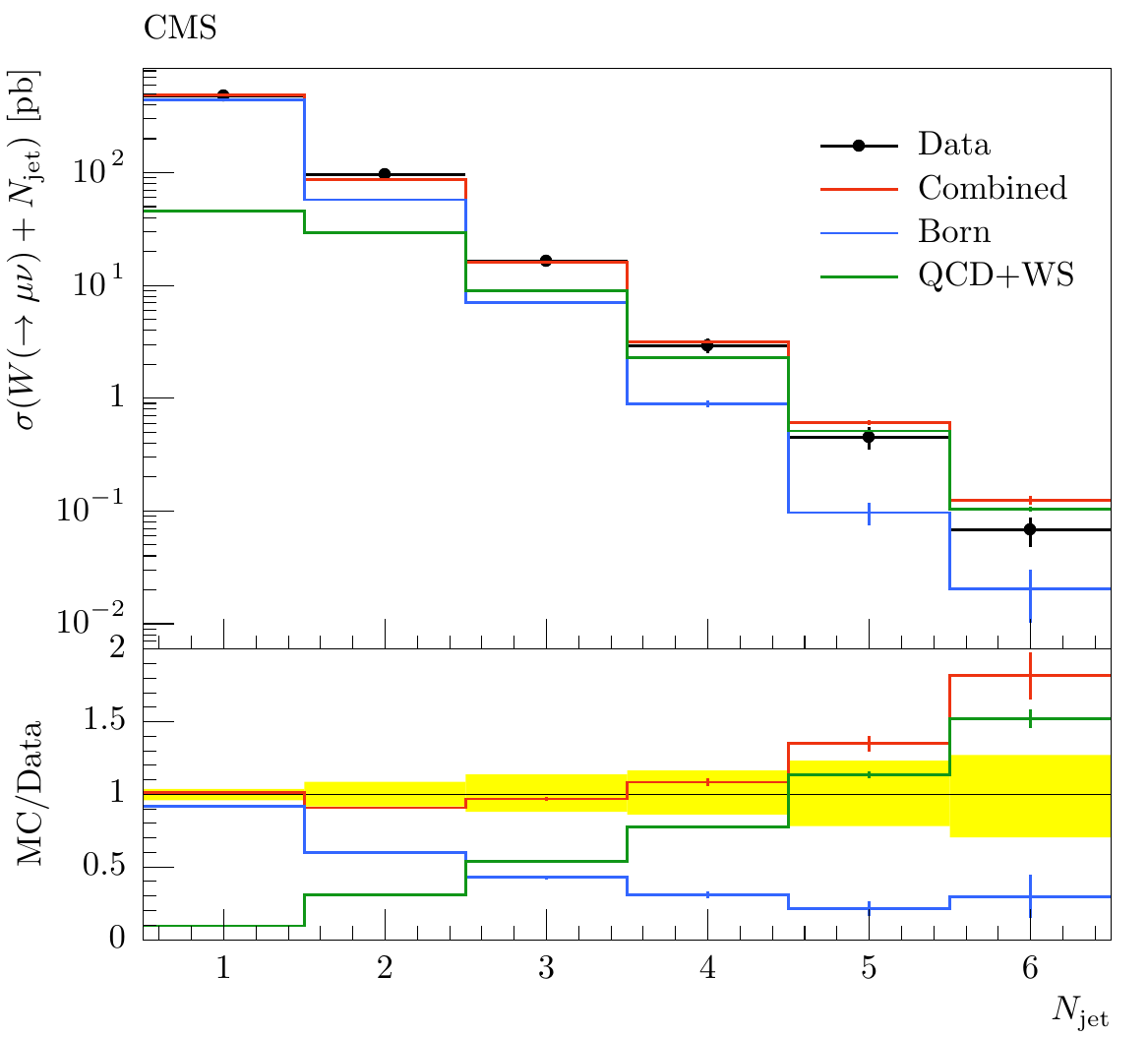}	
		\vspace*{-0.5cm}
	}
\end{minipage}
\end{center}
\vspace{-0.5cm}
\caption{Differential \xsec for (a) $\Z+$jets production and (b) $\W+$jets production in leptonic final states as a function of the number of jets as measured by (a) ATLAS~\cite{ATLASZjets} and (b) CMS~\cite{CMSWjets}. Data points with full uncertainties are compared to the Born description in blue, QCD dijet production with WS in green, and the sum of the two contributions in red. Simulation uncertainties are statistical only.}
\label{fig:jetMultiplicity}
\end{figure}

\begin{figure}[!p]
\begin{center}
\begin{minipage}{0.5\textwidth}
	\centering
	\subfigure[]{
		\includegraphics[width=0.95\linewidth]{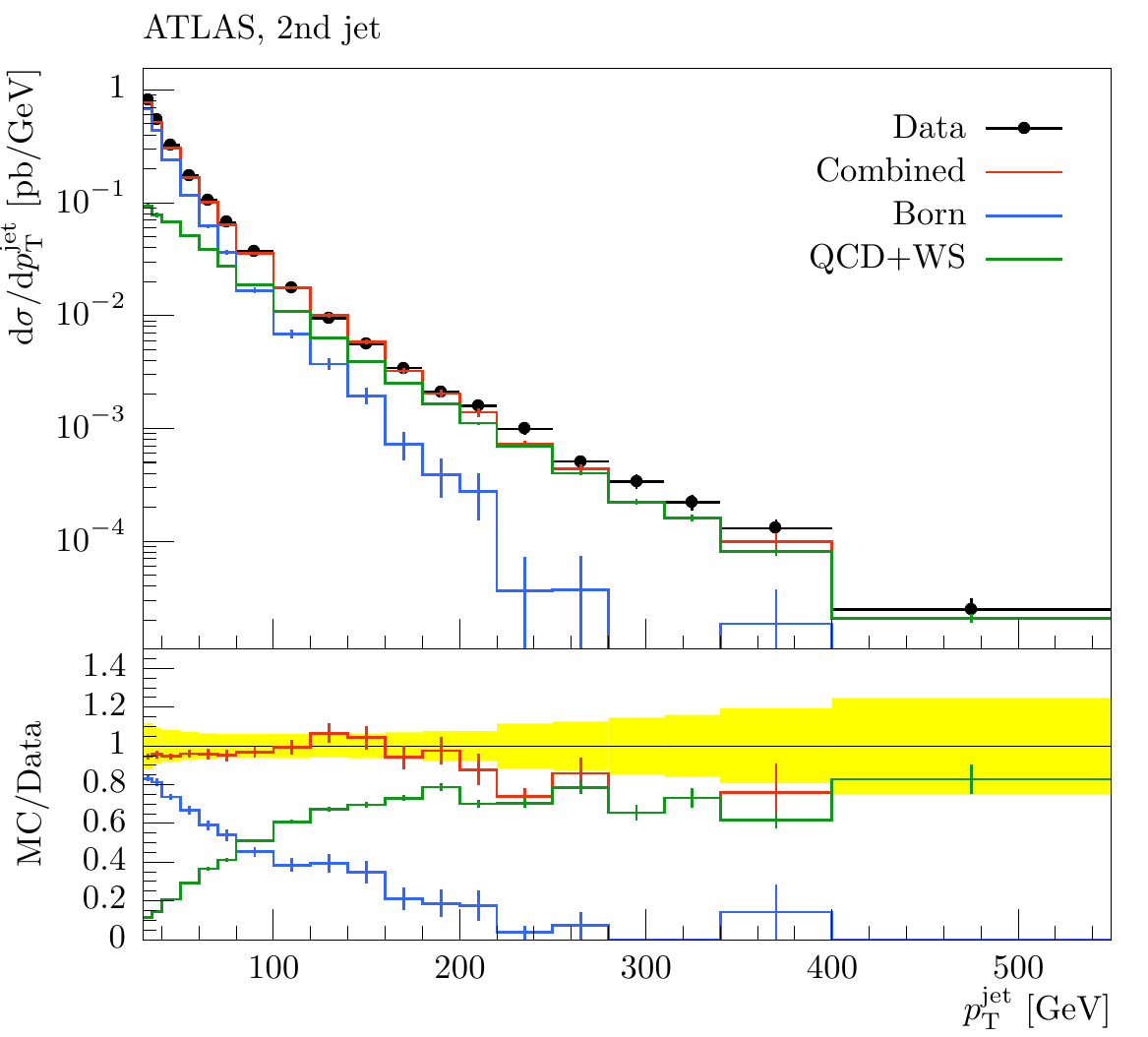}
		\vspace*{-0.5cm}
	}
\end{minipage}
\hspace{-0.25cm}
\begin{minipage}{0.5\textwidth}
	\centering
	\subfigure[]{
		\includegraphics[width=0.95\linewidth]{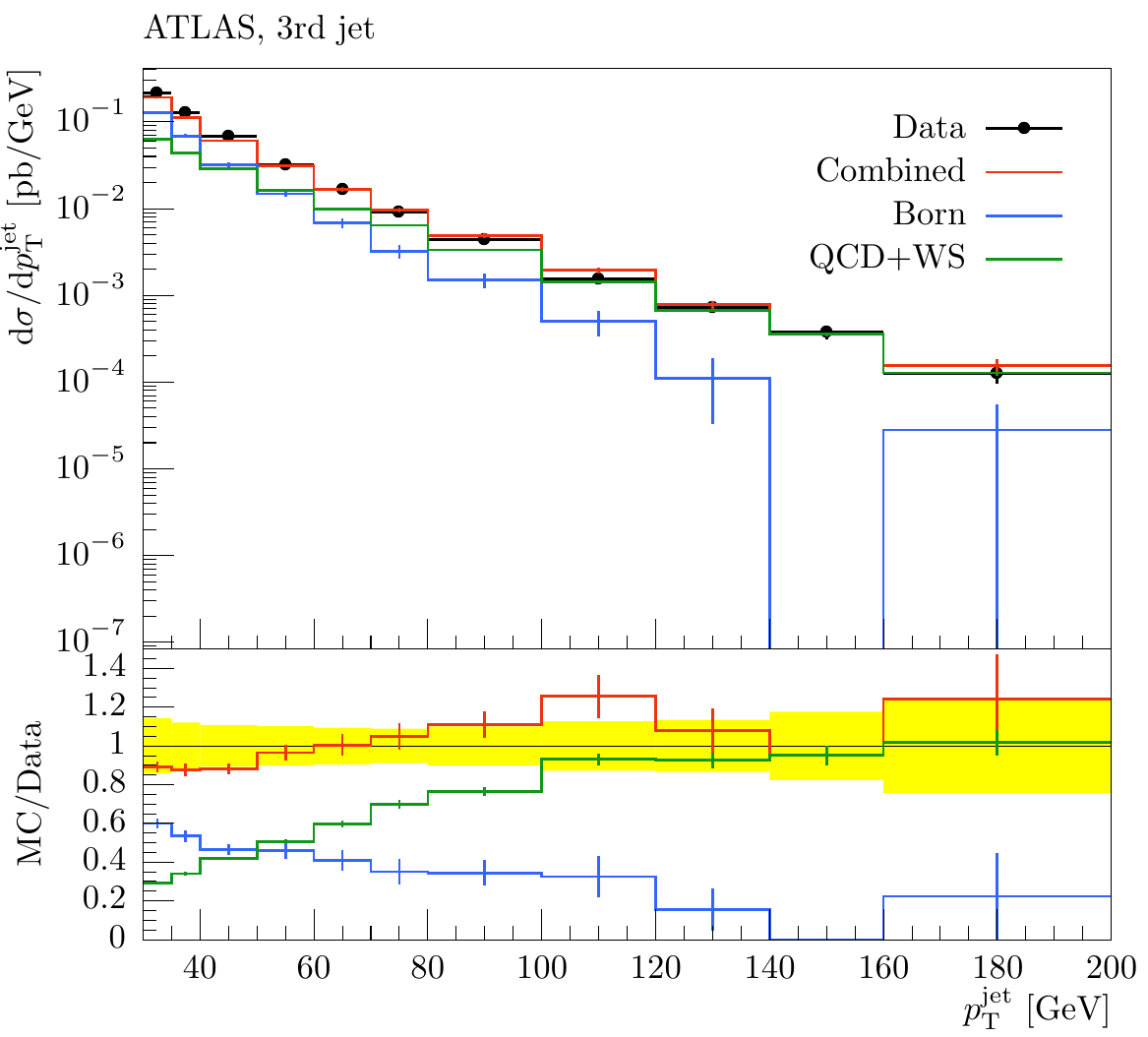}	
		\vspace*{-0.5cm}
	}
\end{minipage}
\begin{minipage}{0.5\textwidth}
	\centering
	\subfigure[]{
		\includegraphics[width=0.95\linewidth]{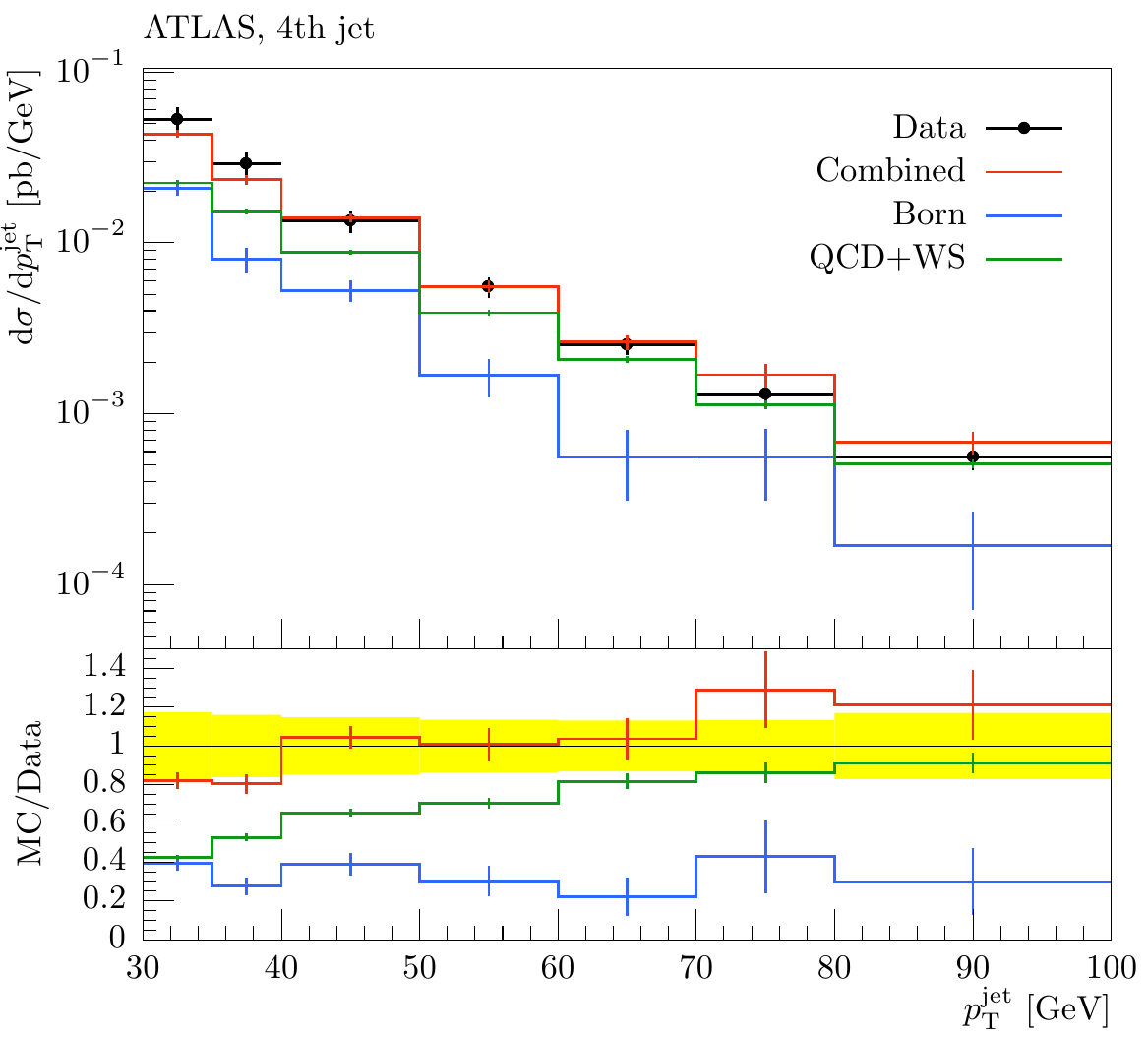}	
		\vspace*{-0.5cm}
	}
\end{minipage}
\end{center}
\caption{Differential \xsec for $\Z+$jets production in leptonic final states as a function of \jetpt for subleading jets as measured by ATLAS~\cite{ATLASZjets}. Data points with full uncertainties are compared to the Born description in blue, QCD dijet production with WS in green, and the sum of the two contributions in red. Simulation uncertainties are statistical only.}
\label{fig:jetpT2to4ATLAS}
\end{figure}

\begin{figure}[tb]
\begin{center}
\begin{minipage}{0.5\textwidth}
	\centering
	\subfigure[]{
		\includegraphics[width=0.95\linewidth]{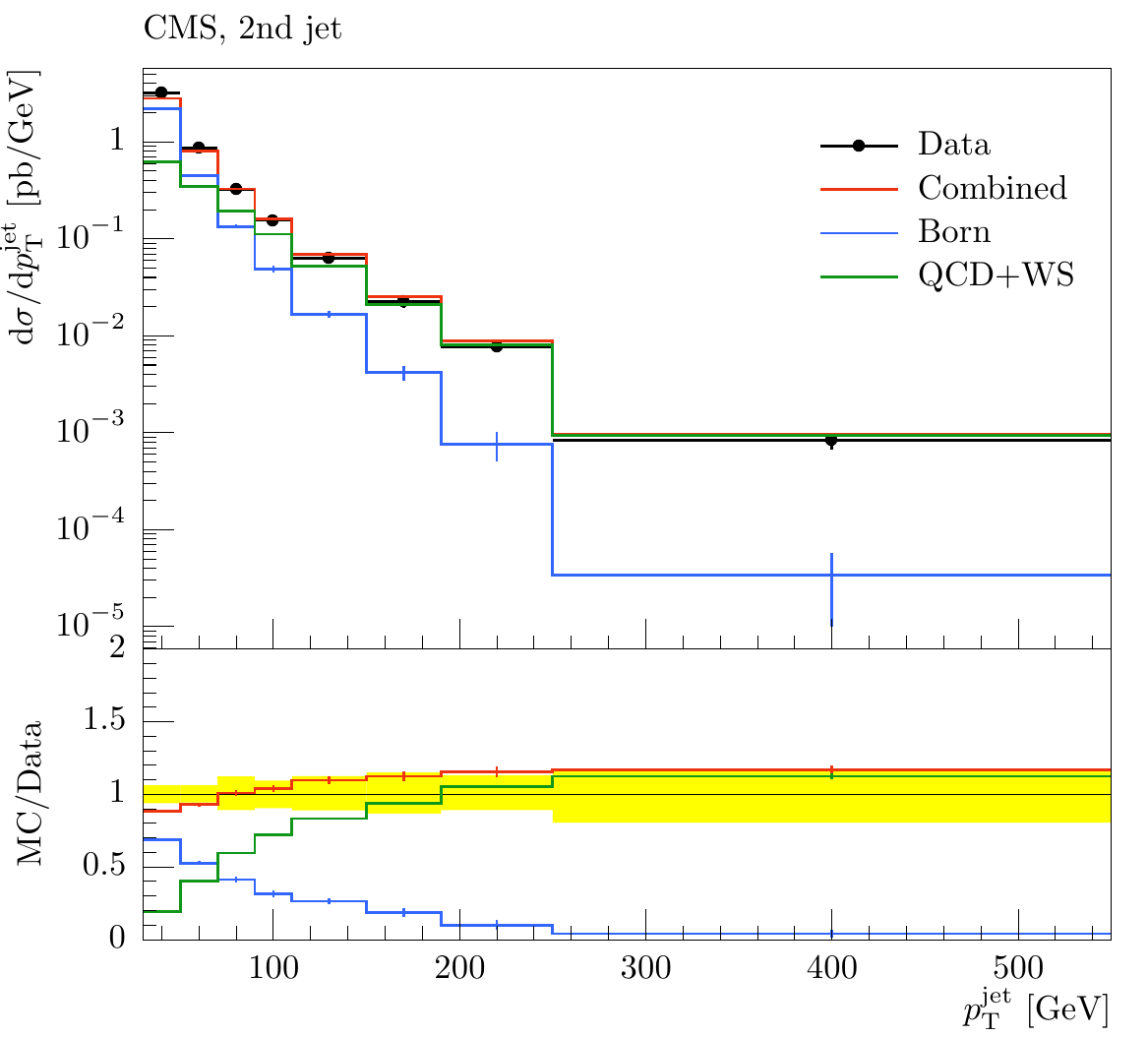}
		\vspace*{-0.5cm}
	}
\end{minipage}
\hspace{-0.25cm}
\begin{minipage}{0.5\textwidth}
	\centering
	\subfigure[]{
		\includegraphics[width=0.95\linewidth]{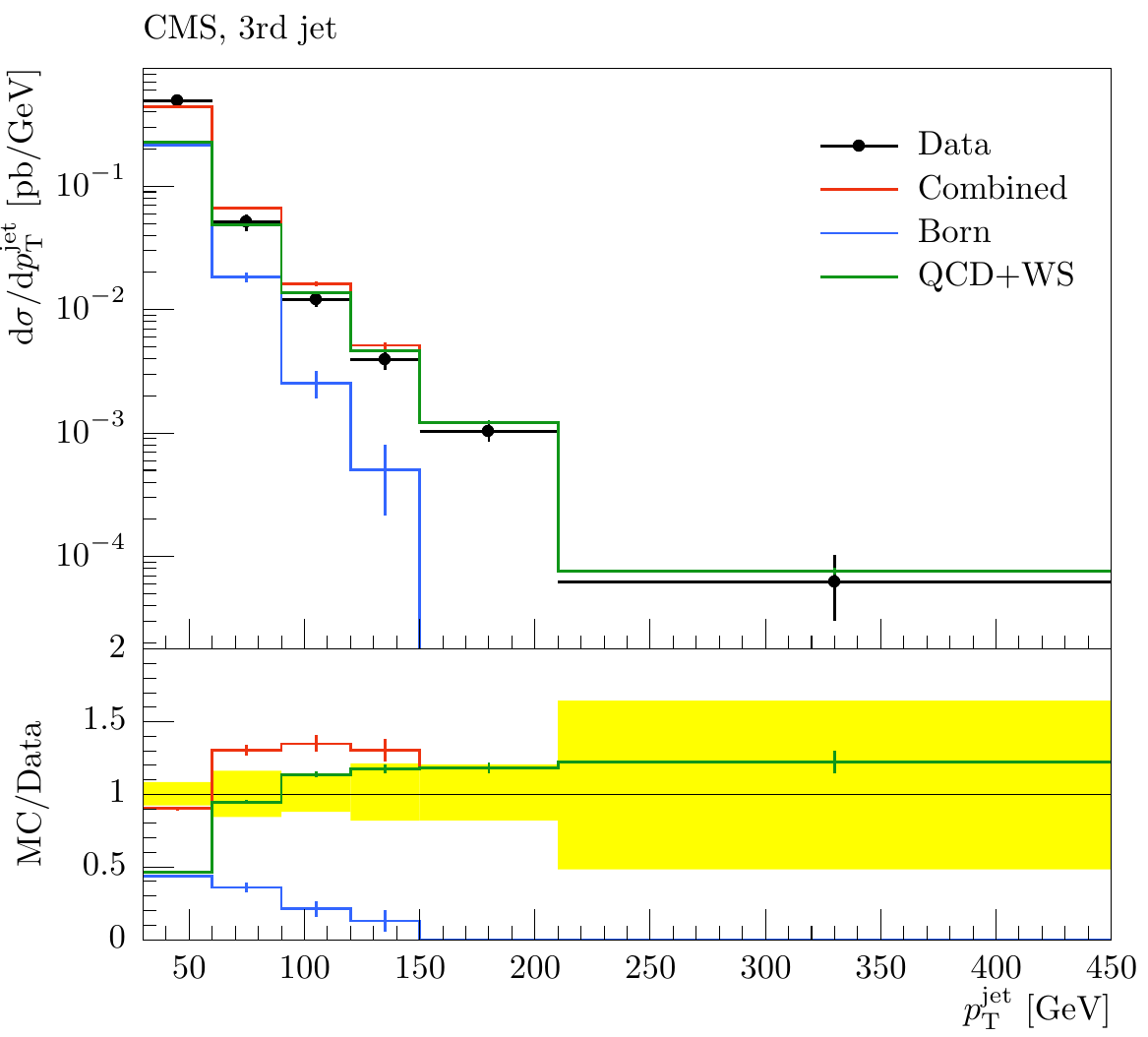}	
		\vspace*{-0.5cm}
	}
\end{minipage}
\begin{minipage}{0.5\textwidth}
	\centering
	\subfigure[]{
		\includegraphics[width=0.95\linewidth]{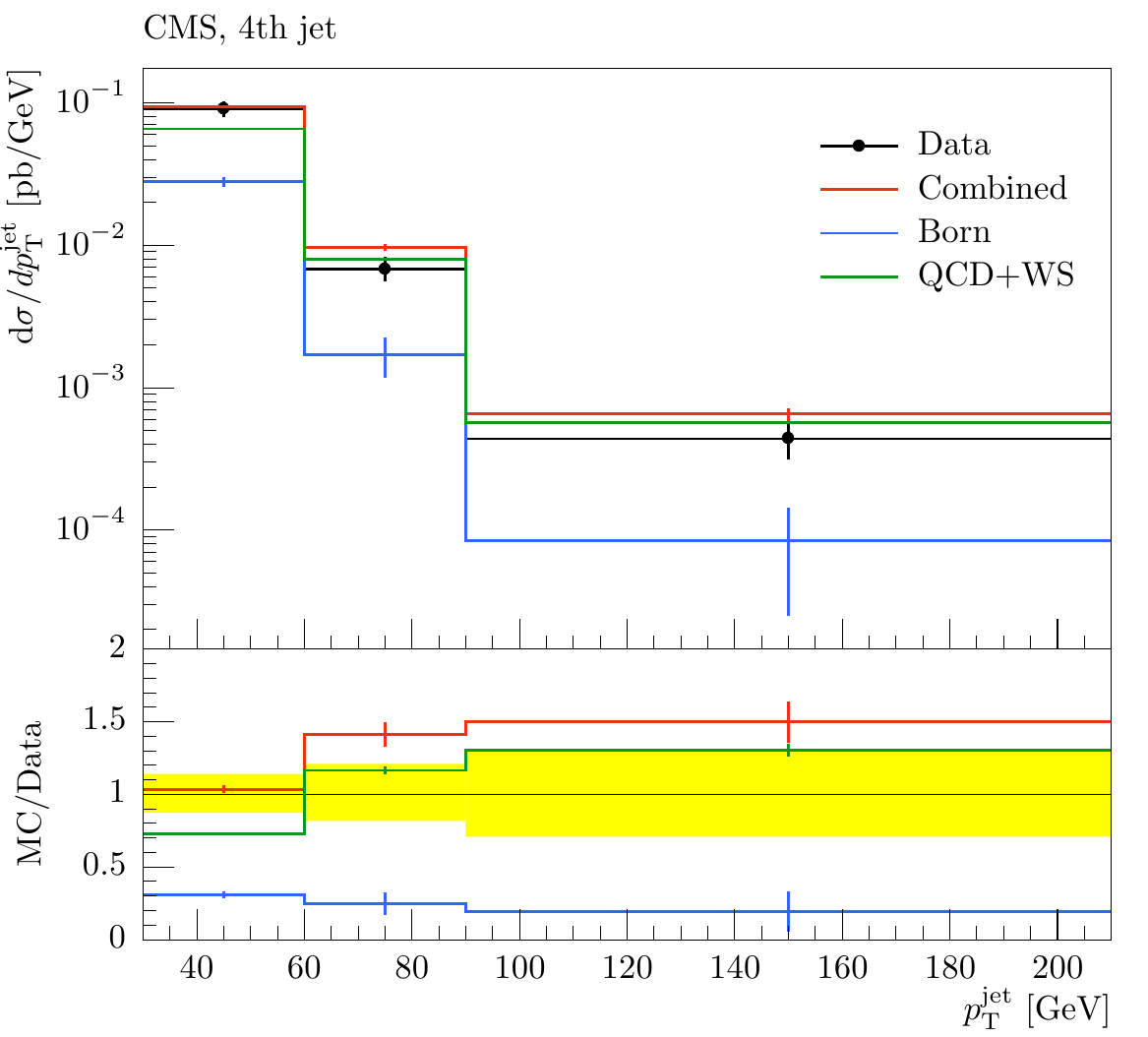}	
		\vspace*{-0.5cm}
	}
\end{minipage}
\end{center}
\caption{Differential \xsec for $\W+$jets production in leptonic final states as a function of \jetpt for subleading jets as measured by CMS~\cite{CMSWjets}. Data points with full uncertainties are compared to the Born description in blue, QCD dijet production with WS in green, and the sum of the two contributions in red. Simulation uncertainties are statistical only.}
\label{fig:jetpT2to4CMS}
\end{figure}

\begin{figure}[tb]
\begin{center}
\begin{minipage}{.5\textwidth}
	\centering
	\subfigure[]{
		\includegraphics[width=0.95\linewidth]{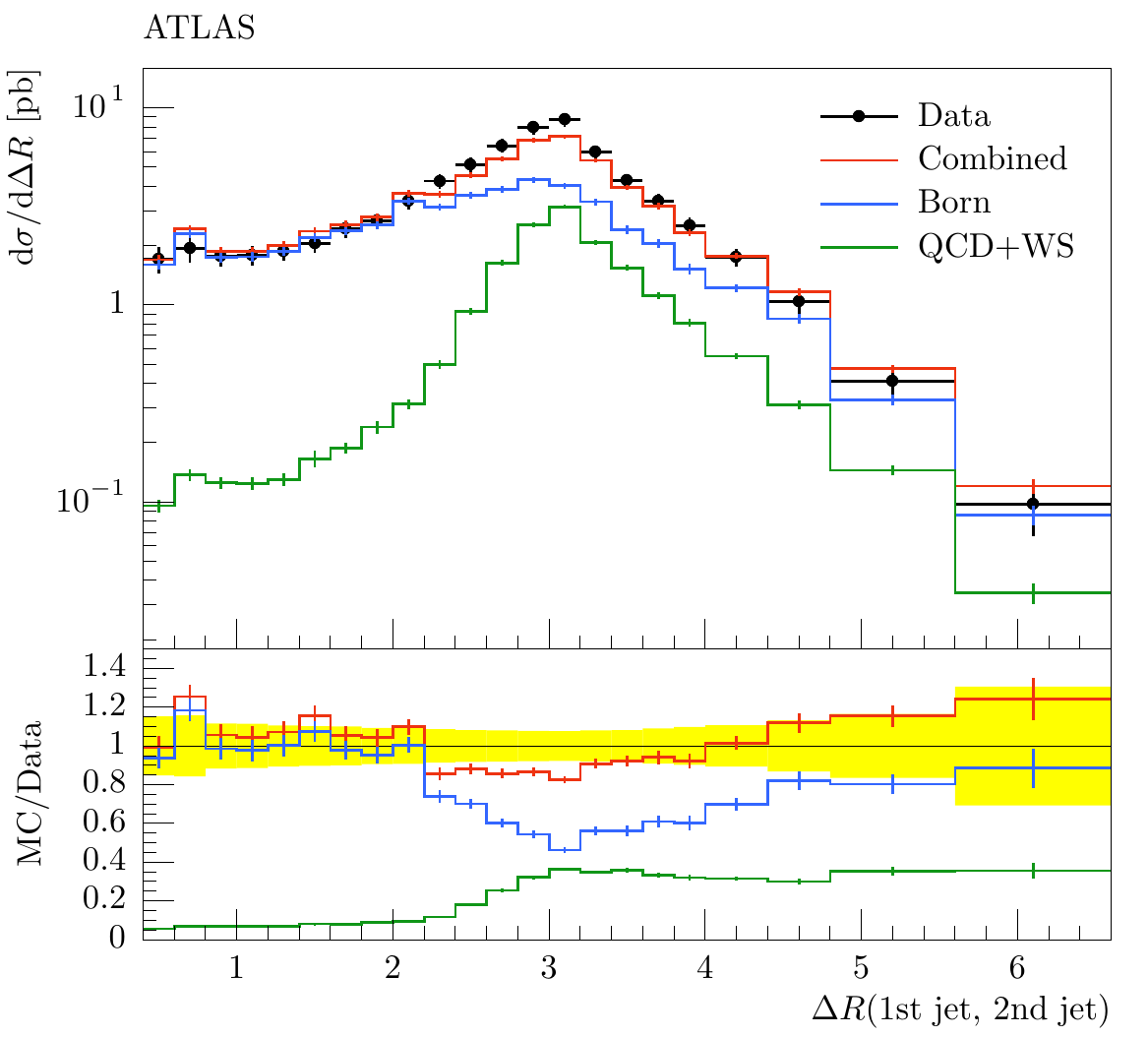}
		\vspace*{-0.5cm}
	}
\end{minipage}
\hspace{-0.25cm}
\begin{minipage}{0.5\textwidth}
	\centering
	\subfigure[]{
		\includegraphics[width=0.95\linewidth]{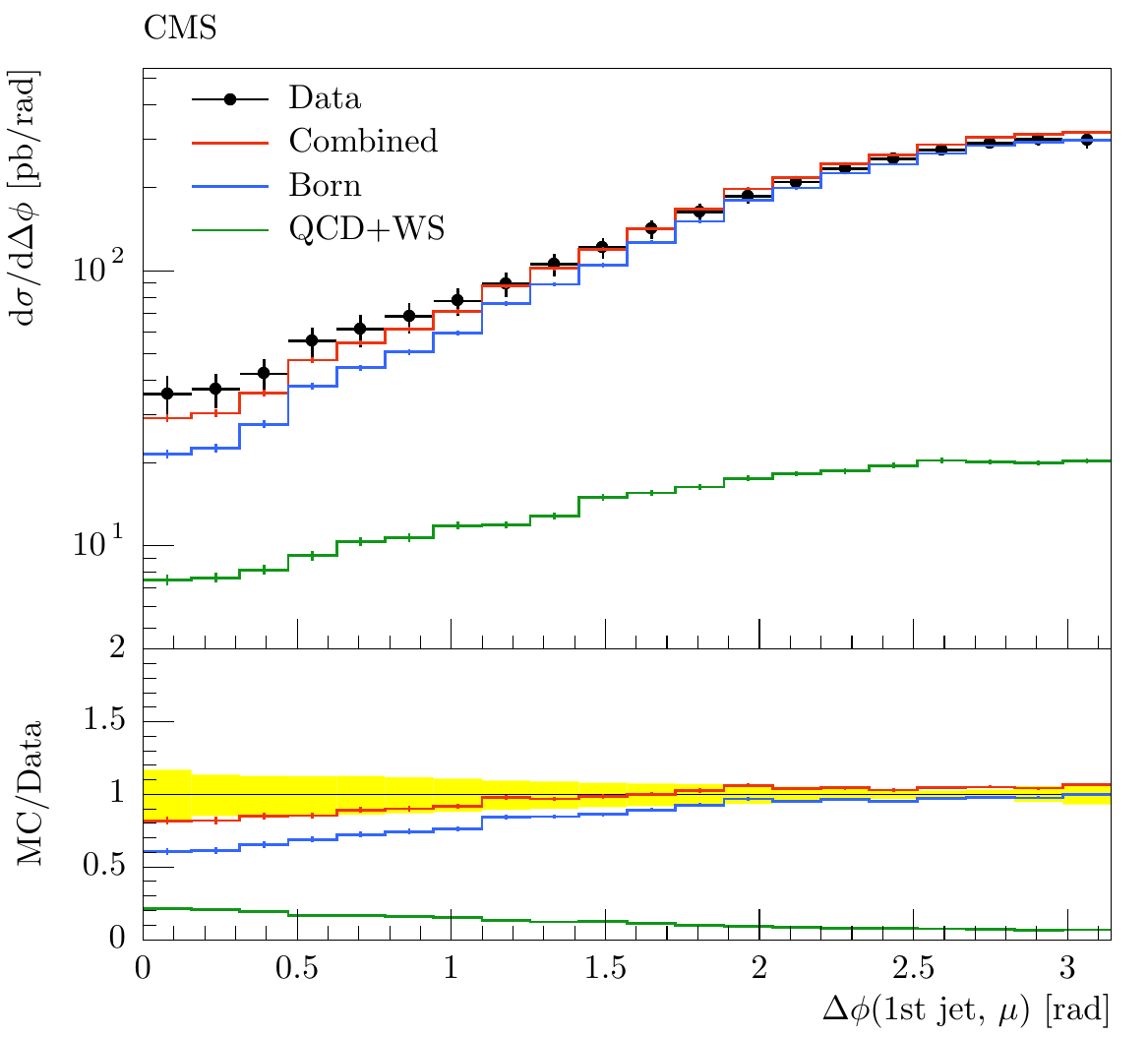}	
		\vspace*{-0.5cm}
	}
\end{minipage}
\end{center}
\vspace{-0.5cm}
\caption{Differential \xsec for (a) $\Z+$jets production as a function of the radial distance between jets as measured by ATLAS~\cite{ATLASZjets} and (b) $\W+$jets production as a function of the azimuthal angle between the muon and the leading jet as measured by CMS~\cite{CMSWjets}. Data points with full uncertainties are compared to the Born description in blue, QCD dijet production with WS in green, and the sum of the two contributions in red. Simulation uncertainties are statistical only.}
\label{fig:eventshapesGPD}
\end{figure}

The comparisons in Figs.~\ref{fig:jetMultiplicity}-\ref{fig:eventshapesGPD} demonstrate that a significant improvement in the modelling of the data is achieved with the inclusion of WS. The number of events with a high jet multiplicity is increased and a better description of \jetpt is achieved. What is more, Figs.~\ref{fig:jetpT2to4ATLAS} and~\ref{fig:jetpT2to4CMS} demonstrate that the weak shower contribution becomes increasingly dominant for subleading jets.

The description of event shapes, while not perfect, is also improved as seen in Fig.~\ref{fig:eventshapesGPD}. This is to be expected from a LO description.

\begin{figure}[tb]
\begin{center}
\begin{minipage}{.5\textwidth}
	\centering
	\subfigure[]{
		\includegraphics[width=0.95\linewidth]{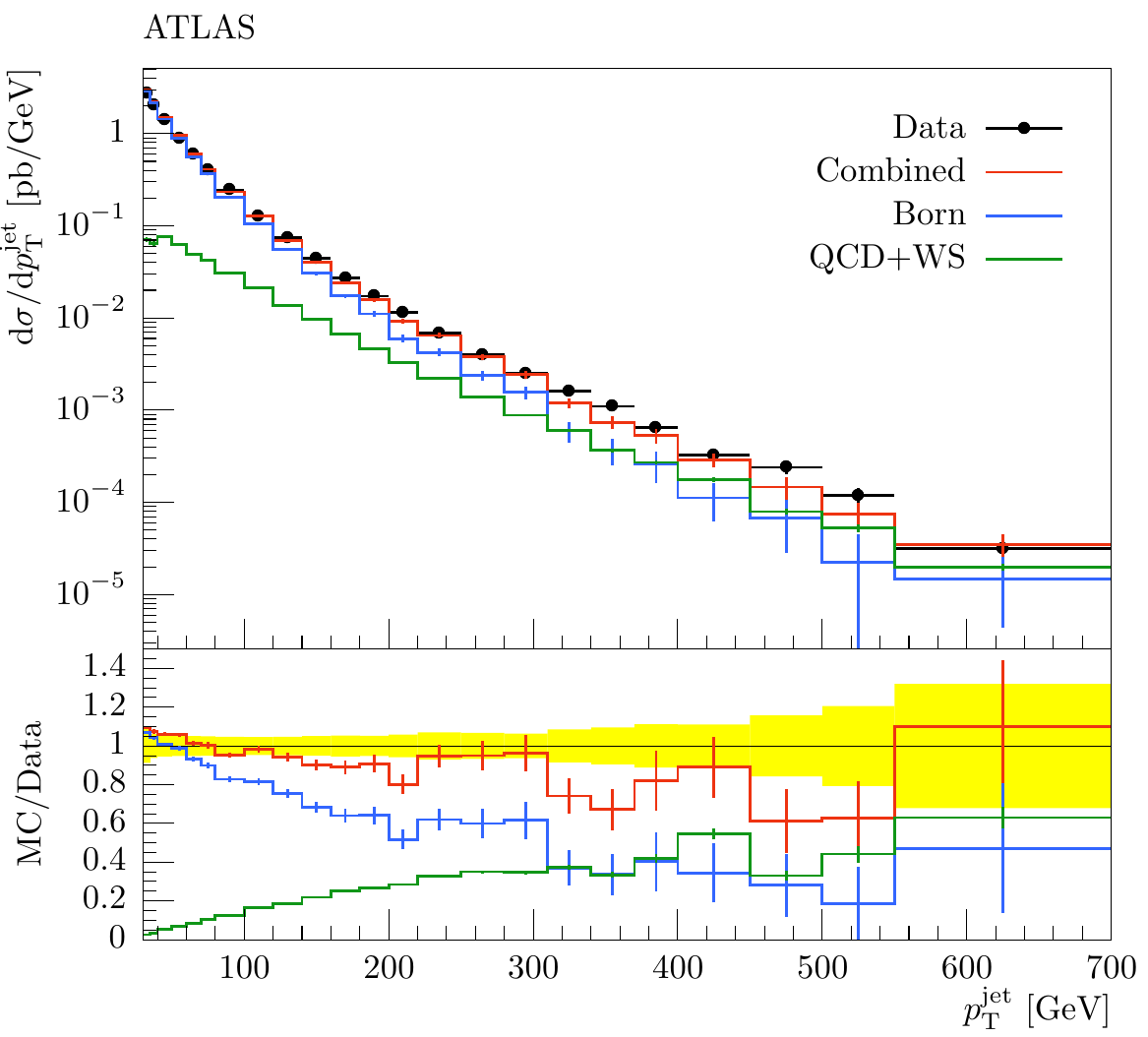}
		\vspace*{-0.5cm}
	}
\end{minipage}
\hspace{-0.25cm}
\begin{minipage}{0.5\textwidth}
	\centering
	\subfigure[]{
		\includegraphics[width=0.95\linewidth]{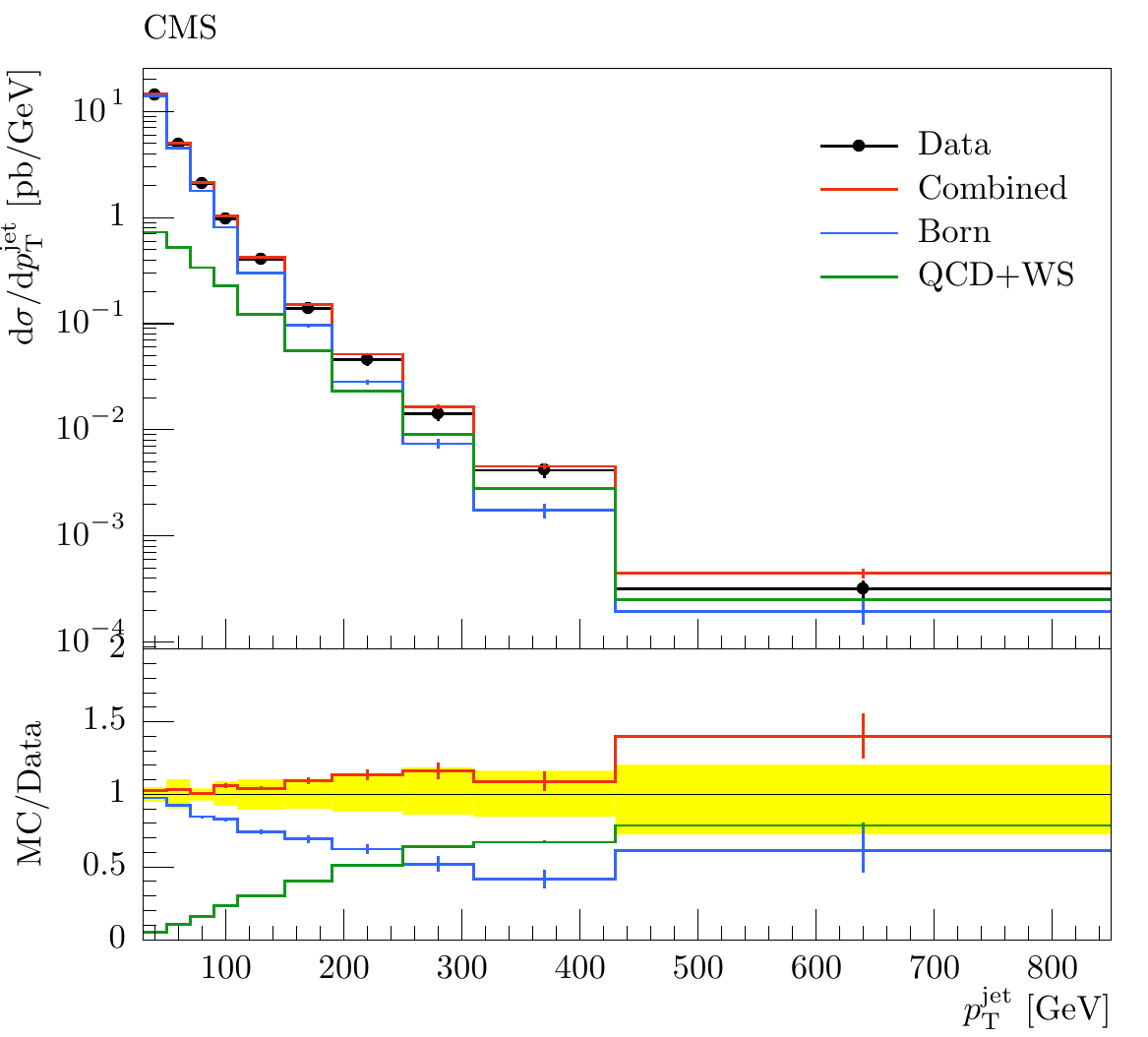}	
		\vspace*{-0.5cm}
	}
\end{minipage}
\end{center}
\vspace{-0.5cm}
\caption{Differential \xsec for (a) $\Z+$jets production and (b) $\W+$jets production in leptonic final states as a function of the transverse momentum of the leading jet as measured by (a) ATLAS~\cite{ATLASZjets} and (b) CMS~\cite{CMSWjets}. Data points with full uncertainties are compared to the Born description in blue, QCD dijet production with WS in green, and the sum of the two contributions in red. Simulation uncertainties are statistical only.}
\label{fig:jetpTGPD}
\end{figure}

Figs.~\ref{fig:jetMultiplicity} and \ref{fig:jetpTGPD} demonstrate that the high \njet and high \jetpt regions, respectively, is overestimated by the combined prediction. This can be attributed to the use of power showers and a value for the strong coupling, $\alpha_s^{\textup{PS}}$, that is significantly higher than the best-fit NLO value at the \Z mass pole, $\alpha_s^{\textup{NLO}}$. Together these serve to compensate for missing higher orders in the calculations. However, the former involves a stretching of the PS beyond its formal region of validity and the effect of the latter grows as $\left(\alpha_s^{\textup{PS}}/\alpha_s^{\textup{NLO}}\right)^{\njet}$. It is, therefore, not surprising that the modelling of high \njet events is poor, in particular.

\begin{figure}[tb]
\begin{center}
\begin{minipage}{.5\textwidth}
	\centering
	\subfigure[]{
		\includegraphics[width=0.95\linewidth]{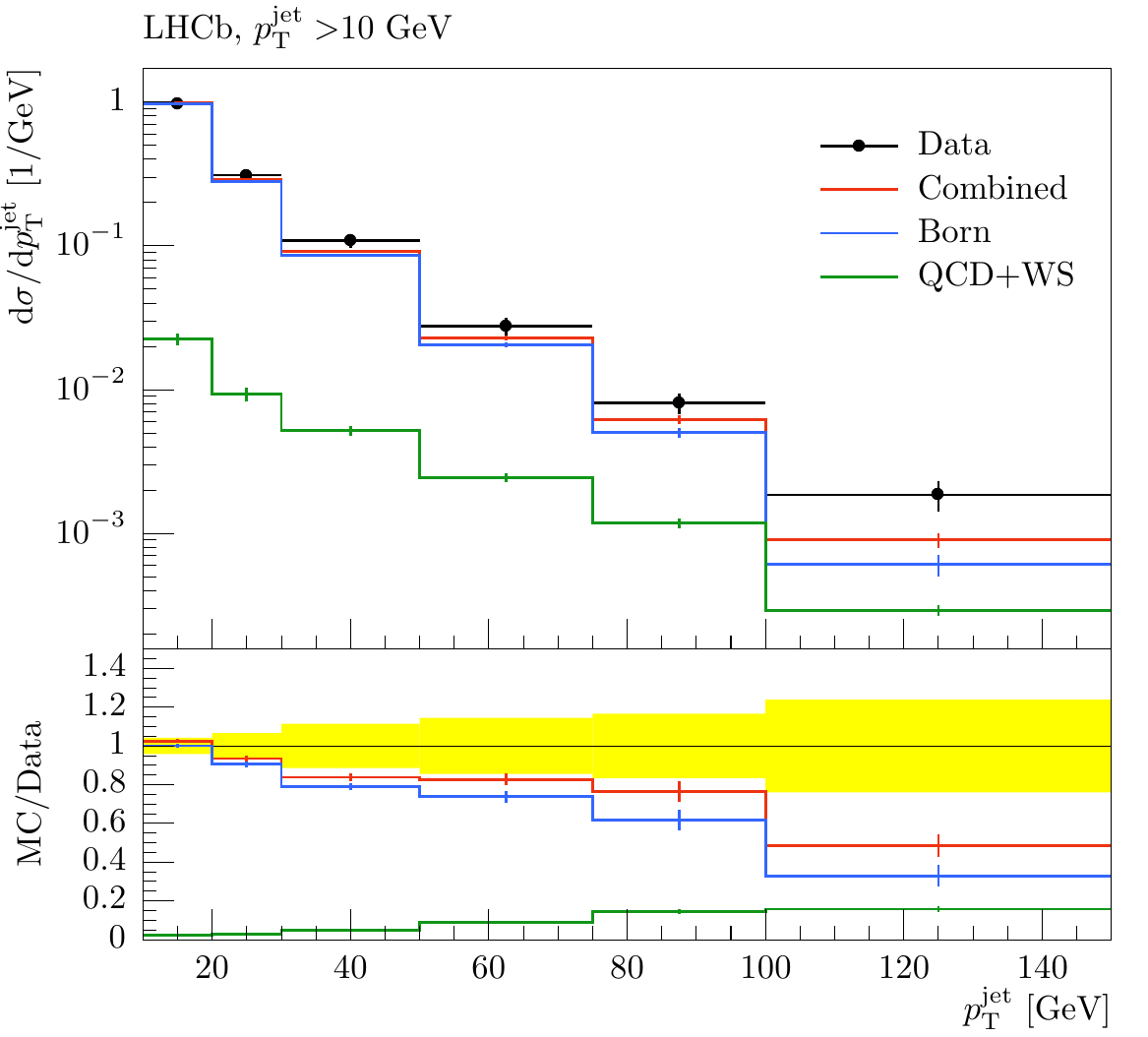}
		\vspace*{-0.5cm}
	}
\end{minipage}
\hspace{-0.25cm}
\begin{minipage}{0.5\textwidth}
	\centering
	\subfigure[]{
		\includegraphics[width=0.95\linewidth]{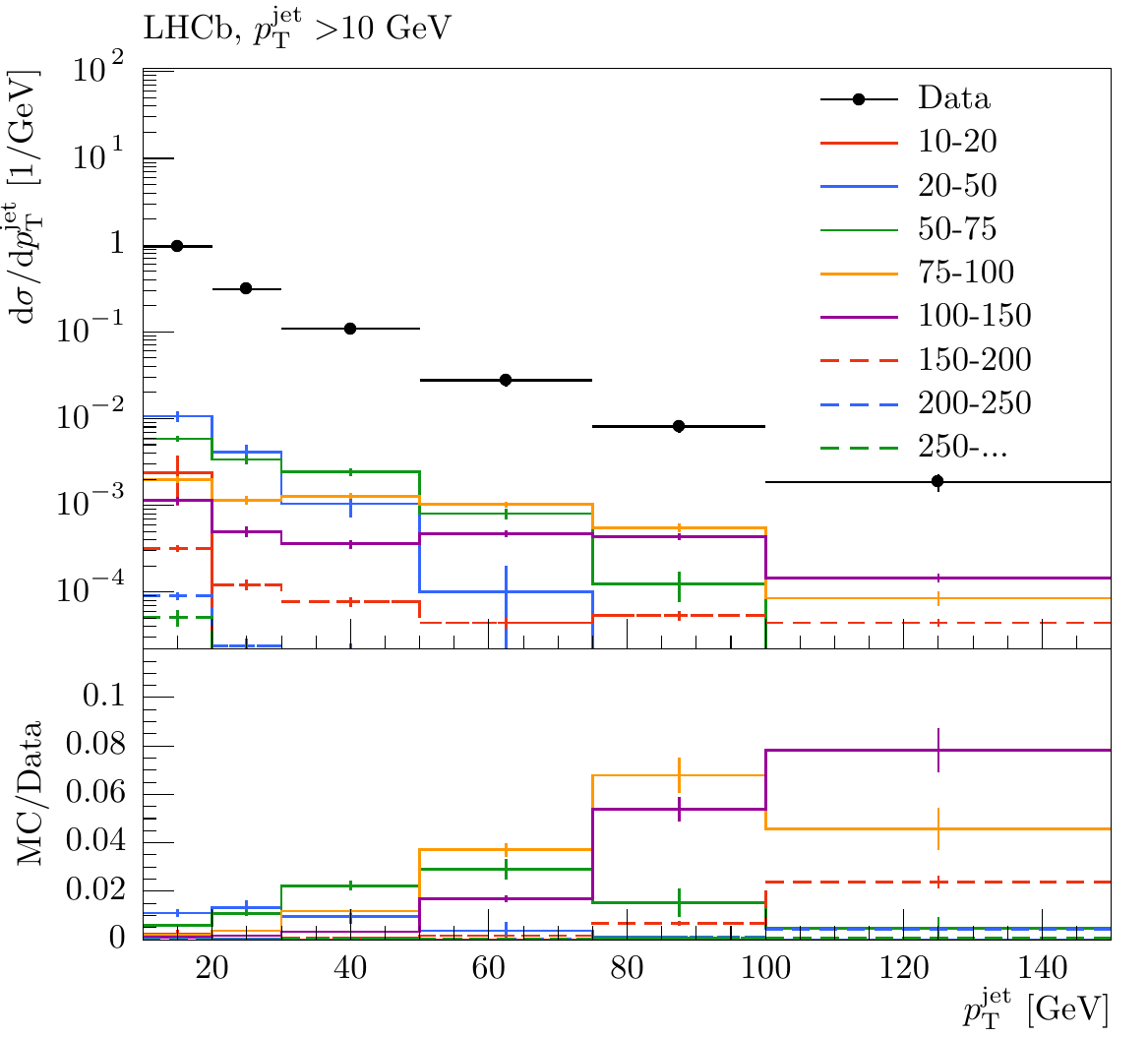}
		\vspace*{-0.5cm}
	}
\end{minipage}
\end{center}
\vspace{-0.5cm}
\caption{Differential \xsec for $\Z+$jets production in the muon final state as a function of \jetpt as measured by LHCb~\cite{LHCbZjets}. In (a) Data points with full uncertainties are compared to the Born description in blue, QCD dijet production with WS in green, and the sum of the two contributions in red. In (b) individual contributions from ranges of outgoing parton transverse momenta in the centre-of-mass frame in dijet production are illustrated. Simulation uncertainties are statistical only.}
\label{fig:jetpTLHCb}
\end{figure}
\begin{figure}[tb]
\begin{center}
\begin{minipage}{.5\textwidth}
	\centering
	\subfigure[]{
		\includegraphics[width=0.95\linewidth]{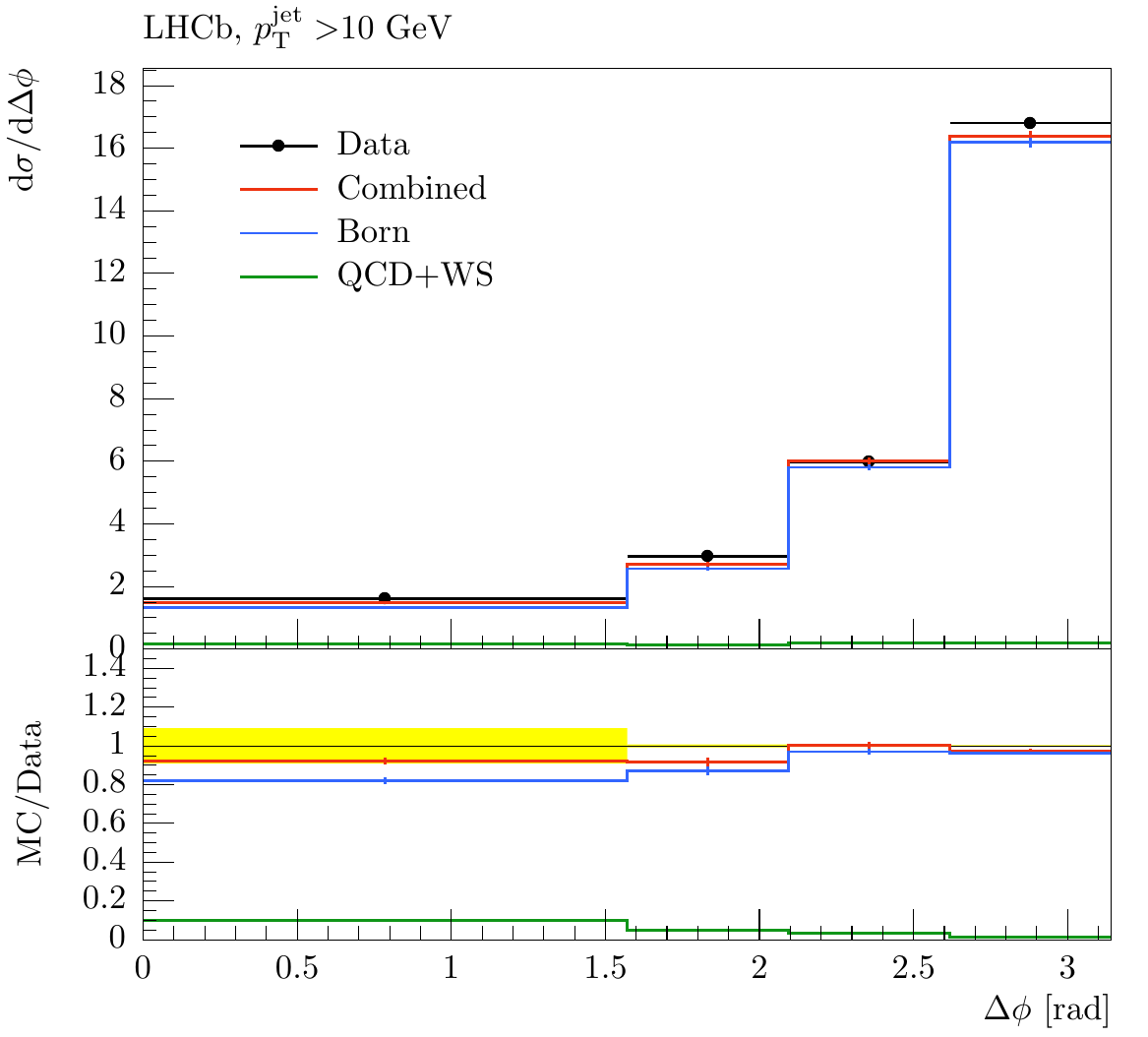}
		\vspace*{-0.5cm}
	}
\end{minipage}
\hspace{-0.25cm}
\begin{minipage}{0.5\textwidth}
	\centering
	\subfigure[]{
		\includegraphics[width=0.95\linewidth]{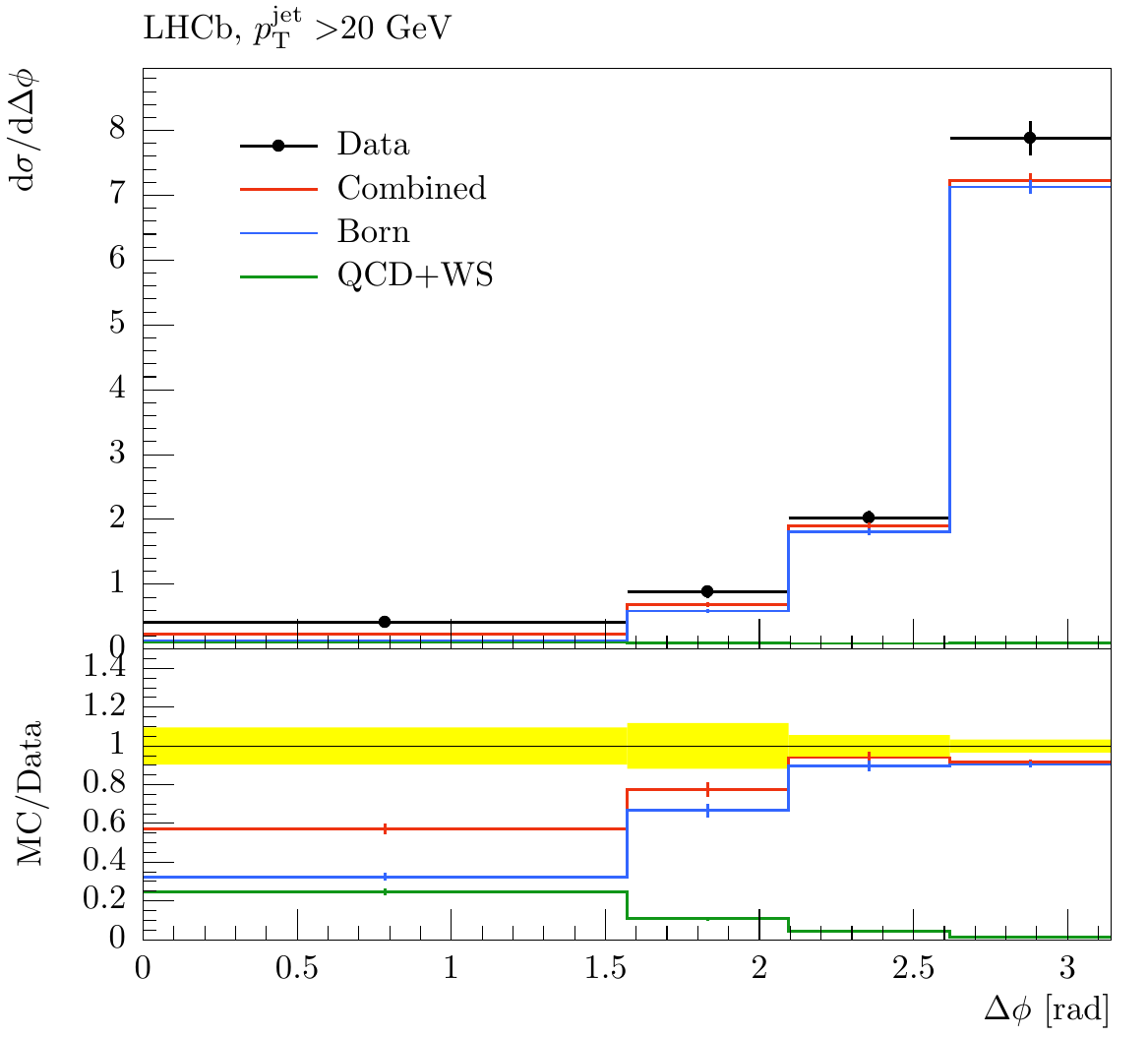}	
		\vspace*{-0.5cm}
	}
\end{minipage}
\end{center}
\vspace{-0.5cm}
\caption{Differential \xsec for $\Z+$jets production in the muon final state as a function of the azimuthal angle between the \Z and the leading jet for (a) $\jetpt>10\gev$ and (b) $\jetpt>20\gev$ as measured by LHCb~\cite{LHCbZjets}. Data points with full uncertainties are compared to the Born description in blue, QCD dijet production with WS in green, and the sum of the two contributions in red. Simulation uncertainties are statistical only.}
\label{fig:dphiLHCb}
\end{figure}

It is evident that LHCb does not yet probe regions of phase space where WS becomes a significant contribution as is illustrated in Figs.~\ref{fig:jetpTLHCb} and \ref{fig:dphiLHCb}. The Born description does not fully account for the data at high \jetpt and low $\Delta \phi$, but neither does the WS component significantly improve agreement with data. Overall, a good description of the LHCb measurement is not achieved in this case.

As a side note, Fig.~\ref{fig:jetpTLHCb} (b) shows individual contributions to the weak shower from ranges of outgoing partonic \pt in the centre-of-mass frame of the underlying $2\to 2$ process. The figure demonstrates that relatively small values of parton transverse momentum can lead to significant subleading contributions to the WS. Specifically, the $[10-20]\gev$ range is the third most important contribution to the first bin of \jetpt. This fact is surprising given that \W and \Z are significantly heavier than the scale of the process and very little phase space is expected to be available for the emission of a massive gauge boson. Emissions are made possible, however, due to the interplay between the large dijet production \xsec at these values of \pt and the fact that \pythia treats the phase space of the event holistically.

\begin{figure}[tb]
\begin{center}
	\includegraphics[width=0.55\linewidth]{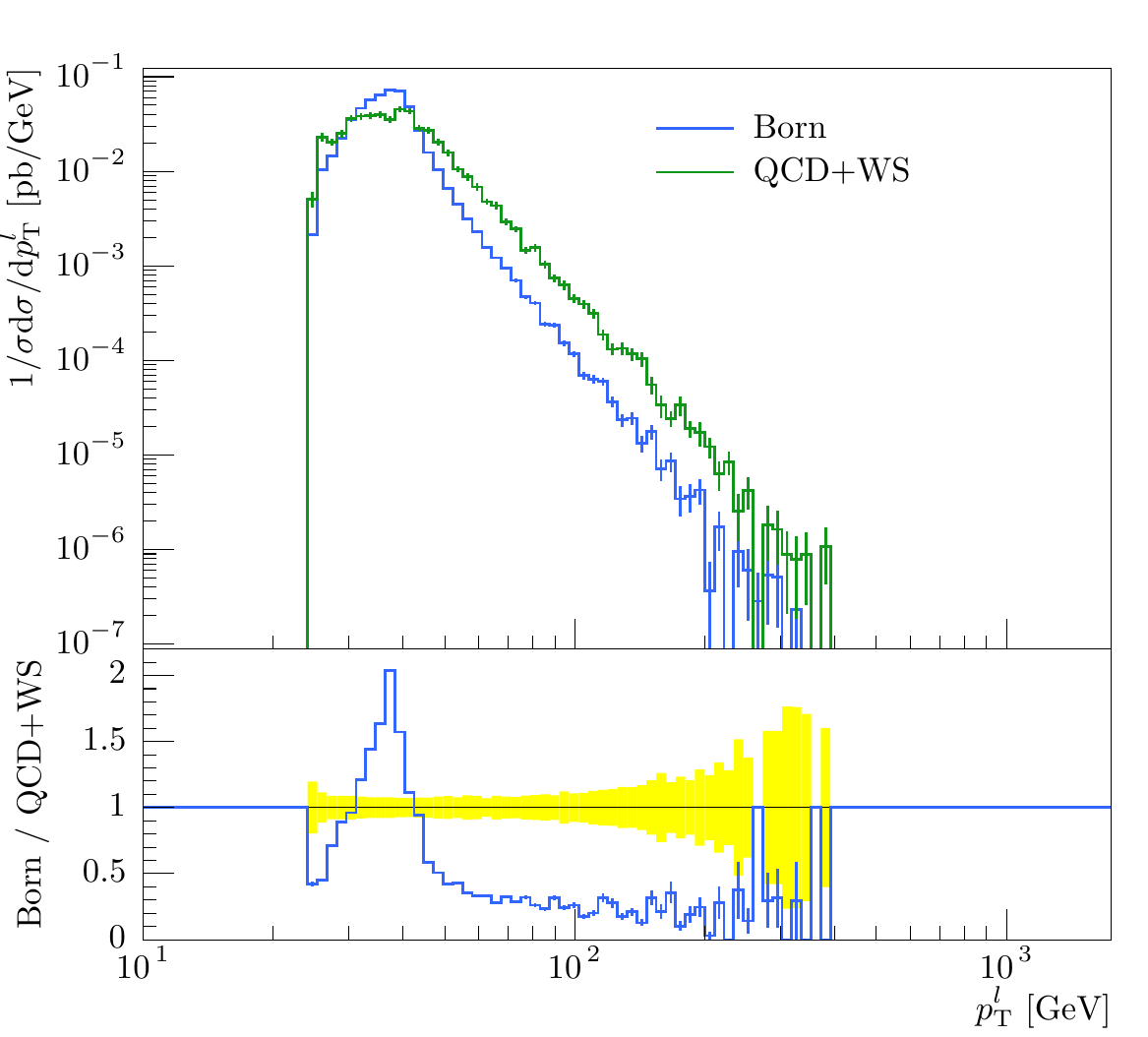}
\vspace*{-0.5cm}
\end{center}
\caption{The lepton transverse momentum distribution in the Born description in blue is compared to QCD dijet production with WS in red. Uncertainties are statistical only and no acceptance cuts have been applied.}
\label{fig:leptonpT}
\end{figure}

Finally, Fig.~\ref{fig:leptonpT} shows the \pt spectra of \W daughters in leptonic decays, \ptl, for the Born and weak shower description. No acceptance cuts have been applied in this case. It is clear that leptons which originate from the PS have a wider Jacobian peak and have the effect of smearing the \ptl distribution and increasing the contribution from high-\ptl events. This impacts EW analyses that use fits to the \ptl distribution to extract physical quantities such as event yields or the \W boson mass.

\clearpage

\section{Conclusions}
\label{sec:conclusions}

This note presents a study of \mnm schemes and WS. These novel advanced theoretical tools are compared to measurements performed at the LHC by the ATLAS, CMS, and \lhcb collaborations at $\sqrt{s}=7$\tev and $\sqrt{s}=13$\tev.

\mnm schemes are seen to provide a good description of event shapes in the central and forward regions and to model \jetpt and \ptZ distributions accurately in the forward region. WS, while accurate to LO only, is able to significantly improve the description of the jet multiplicity and \jetpt distributions centrally. WS also yields an improvement in the modelling of event shapes. Finally, measurements performed at \lhcb do not yet access regions of phase space where WS is expected to be dominant. WS becomes a significant, or even the dominant, production mechanism of EW bosons for event topologies with two, or more, jets and scales beyond 100\gev in \jetpt. It is clear, however, that with increased data at higher centre-of-mass energies will improve the reach of measurements performed at \lhcb. In parallel, WS effects such as the suppression of the jet rate as well as multiple emissions are expected to become more common at these energies.

This study constitutes a test of precise theoretical tools in a novel region of phase space. It proves that a variety of \mnm schemes can be employed with some confidence in modelling EW boson production in association with jets. Finally, it is clear that with the extraordinary success of the SM and with beyond-the-Standard-Model physics proving elusive, increased precision is required in both the experimental measurements' programme at the LHC as well as in developing theoretical tools such as matching and merging and weak showering.

\section{Acknowledgements}
\label{sec:acknowledgements}

We take this opportunity to thank Leif L{\"o}nnblad and Torbj{\"o}rn Sj{\"o}strand for their expert guidance during the course of this research. We would also like to acknowledge the invaluable help we received from Jesper Roy Christiansen and Stefan Prestel and the entire Lund theory group. Finally, we would like to thank David Ward for improving the quality of this note and for many helpful suggestions. This work was supported in part by the European Union as part of the FP7 Marie Curie Initial Training Network MCnetITN (PITN-GA-2012-315877).

%\newpage
%\input{appendix}

\newpage

\addcontentsline{toc}{section}{References}
\setboolean{inbibliography}{true}
\bibliographystyle{bibStyle}

\bibliography{mine.bib}

\end{fmffile}
\end{document}